\documentclass[aps,prl,floatfix,twocolumn,reprint,amsmath,amssymb,superscriptaddress]{revtex4-1}
\usepackage{amsfonts}
\usepackage{mathrsfs}
\usepackage{amsmath}% needed for subequations
\usepackage{color}
\usepackage{natbib}
\usepackage{graphicx}
\usepackage{bm}% bold maths
\usepackage{amssymb}
\usepackage{xspace}
\usepackage{epstopdf}
\usepackage{dcolumn}% Align table columns on decimal point
\usepackage{longtable}
\usepackage{array}
\usepackage{makecell}
\usepackage{tabulary}
\usepackage{multirow}
\newcolumntype{C}[1]{>{\centering\arraybackslash}p{#1}}
\newcolumntype{L}[1]{>{\flushleft\arraybackslash}p{#1}}

\usepackage{multirow}
\usepackage{epsfig}
\usepackage[normalem]{ulem}
\usepackage{amsmath}
\usepackage{braket}
\usepackage{bbold}
\usepackage{natbib}

\usepackage[colorlinks=true, letterpaper=true, pdfstartview=FitV, linkcolor=blue, citecolor=blue, urlcolor=blue]{hyperref}

\makeatletter

\newcommand{\Rmnum}[1]{\expandafter\@slowromancap\romannumeral #1@}
\makeatother
\begin{document}

\title{Nodal Line Spin-gapless Semimetals and High-quality Candidate Materials}

\author{Run-Wu Zhang}
\thanks{These authors contributed equally to this work.}
\affiliation{Key Lab of advanced optoelectronic quantum architecture and measurement (MOE), Beijing Key Lab of Nanophotonics $\&$ Ultrafine Optoelectronic Systems, and School of Physics, Beijing Institute of Technology, Beijing 100081, China}

\author{Zeying Zhang}
\thanks{These authors contributed equally to this work.}
\affiliation{Key Lab of advanced optoelectronic quantum architecture and measurement (MOE), Beijing Key Lab of Nanophotonics $\&$ Ultrafine Optoelectronic Systems, and School of Physics, Beijing Institute of Technology, Beijing 100081, China}

\author{Cheng-Cheng Liu}
\email{ccliu@bit.edu.cn}
\affiliation{Key Lab of advanced optoelectronic quantum architecture and measurement (MOE), Beijing Key Lab of Nanophotonics $\&$ Ultrafine Optoelectronic Systems, and School of Physics, Beijing Institute of Technology, Beijing 100081, China}

\author{Yugui Yao}
\email{ygyao@bit.edu.cn}
\affiliation{Key Lab of advanced optoelectronic quantum architecture and measurement (MOE), Beijing Key Lab of Nanophotonics $\&$ Ultrafine Optoelectronic Systems, and School of Physics, Beijing Institute of Technology, Beijing 100081, China}

%\date{\today}
%
\begin{abstract}
Spin-gapless semimetals (SGSMs), which generate 100\% spin polarization, are viewed as promising semi-half-metals in spintronics with high speed and low consumption. We propose and characterize a new $\mathbb{Z_{\mathrm{2}}}$ class of topological nodal line (TNL) in SGSMs. The proposed TNLSGSMs are protected by space-time inversion symmetry or glide mirror symmetry with two-dimensional (2D) fully spin-polarized nearly flat surface states. Based on first-principles calculations and effective model analysis, a series of high-quality materials with $\textit{R}\overline{3}\textit{c}$ and $\textit{R}{3}\textit{c}$ space groups are predicted to realize such TNLSGSMs (chainlike). The 2D fully spin-polarized nearly flat surface states may provide a route to achieving equal spin pairing topological superconductivity as well as topological catalysts.

\end{abstract}
\maketitle

%%%%%%%%%%%%%%%%%%%%%%%%%%%%%%%%%%%%%%%
\textit{\textcolor{black}{Introduction.---}} Spintronics, using an electron's spin instead of its charge to carry information and featured high speed and low energy consuming, has attracted tremendous interest from academic research to industrial applications in recent years~\cite{1wolf2001spintronics,2vzutic2004spintronics}. Half-metals, which manifest one spin channel possessing metallic states while the other keeping insulating or semiconducting. Unlike conventional ferromagnetic (FM) alloys with a low degree of spin polarization, half-metals with 100\% spin polarization are regarded as excellent spintronics candidates for spin generation, injection, and transport. Recently, as a remarkable upgraded version of half-metals, spin-gapless semiconductors (SGSs) had been proposed ~\cite{3wang2008proposal}. In addition to the advantages of the standard half-metals, the SGSs have large electron mobility and highly tunable capabilities by external fields, such as pressure, electric fields, magnetic fields, electromagnetic radiation, \textit{etc}. The unique SGSs provide a new playground and opportunities for spintronics, electronics, and optics. Especially, the SGSs own ideal Weyl points for the metallic spin channel without other entangled trivial bands, which marry spintronics and topological physics.

Topological semimetals (TSMs) are systems where the conduction and the valence bands cross each other with robustness in the Brillouin zone (BZ). Among TSMs, topological nodal line semimetals (TNLSs) are regarded as a new class of topological quantum states~\cite{4mullen2015line, 5fang2015topological, 6weng2015topological, 7kim2015dirac, 8yu2015topological, 9ezawa2016loop, 10bian2016topological, 11hu2016evidence, 12schoop2016dirac, 13li2016dirac, 14kobayashi2017crossing, 15sun2017dirac, 16li2017type, 17wang2017antiferromagnetic, 18zhang2018nodal, 19ma2018mirror, 20gong2018symmorphic, 21bzduvsek2016nodal, 22wang2017hourglass, 23yan2017nodal,  24chen2017topological, 25fu2018hourglasslike, 26singh2018topological, 27feng2018topological, 28wang2018topological, 29ezawa2017topological, 30bi2017nodal}, which bridge the gapped and gapless phases, can be driven into various topological phases, such as topological insulators and Dirac (Weyl) semimetals~\cite{31kane2005quantum, 32bernevig2006quantum, 33wan2011topological, 34young2012dirac, 35yang2014classification, 36wang2012dirac, 37wang2013three, 38chang2017type, 39xu2011chern, 40weng2015weyl, 41soluyanov2015type, 42ruan2016ideal, 43autes2016robust, 44zhu2016triple, 45wang2017prediction}. Many intriguing physical properties have been proposed in these interesting TNLSs, including Friedel oscillation,~\cite{13li2016dirac} non-dispersive Landau energy level,~\cite{46rhim2015landau} and specific long-range Coulomb interactions~\cite{47huh2016long}, \textit{etc}.

Buoyed by the aforementioned superior properties, the efficient combination of spin-gapless feature and TNLS nature (termed as TNLSGSM thereafter) is desirable. In this Letter, we first showcase a general classification of TNLSGSMs and further propose an enhanced NL connection mode (dubbed as a nodal chain SGSM). The coexistence of spin-polarized and linear dispersion in the NL states can be realized in a family of real materials with $\textit{R}\overline{3}\textit{c}$ and $\textit{R}{3}\textit{c}$ space groups. These ideal candidates feature ultra-clean energy dispersion and ultra-high Fermi velocity, which vastly enriches the TNLSs and half-metals family. Distinguished from most proposed nodal chain states, the four-fold degeneracy neck crossing-point traces out two-fold degeneracy lines emerged in the single spin channel and the profile of nodal chainlike SGSMs are revealed by the tight-binding (TB) model. The efficient combination of nodal chain and fully spin-polarized nature in feasible materials greatly refresh the thoughts for designing potential high-performance spintronic devices.

\begin{figure}
\includegraphics[width=1.0\columnwidth]{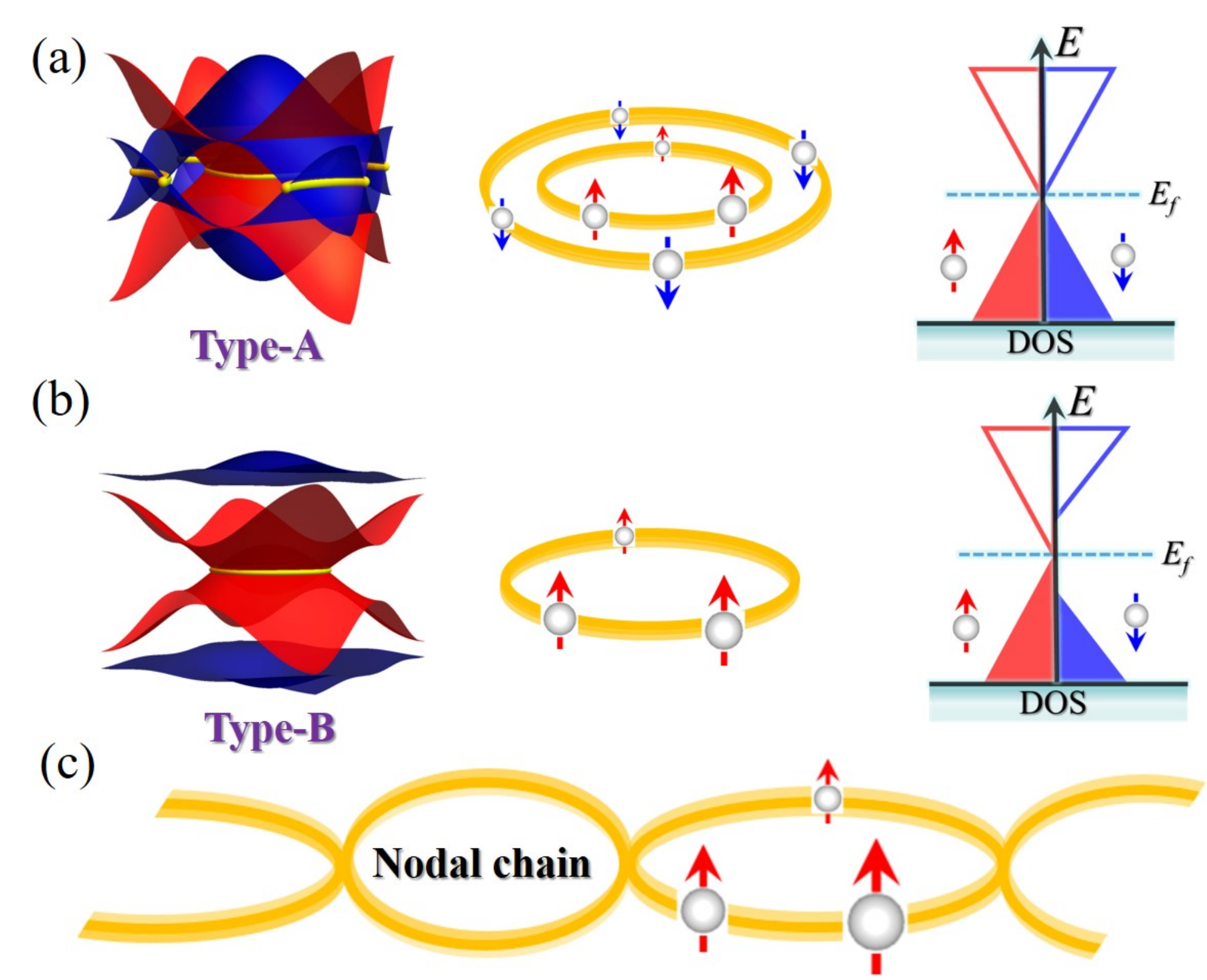}
\caption
{(Color online) Schematic illustration of the simplest TNLSGSMs between two inverted sub-bands and the corresponding density of states (DOS) for (a) concentric loops come from different spin channels separately and (b) NLs come from a single spin channel. (c) Schematic figure showing nodal chain SGSMs consisting of TNLSGSMs. The red (blue) pattern represents the spin-up (spin-down) channel.}
\label{fig1}
\end{figure}

\textit{\textcolor{black}{Nodal Line Spin-gapless Semimetal.---}} According to the different spin-polarized band crossing patterns at the Fermi level, NLSGSMs could give rise to two possible configurations. Here, we construct a two-band model for describing the classifications of NLSGSM. Considering an effective Hamiltonian near $\varGamma$
\begin{eqnarray}
H(k)& = & [(\cos k_{x}+\cos k_{y}-r_{s})\sigma_{1}+\sin k_{z}\sigma_{3}]s_{3}\nonumber \\
 &  & +m(1-s_{3})\sigma_{2},
\end{eqnarray}
where Pauli matrices $\sigma_{i=1,2,3}$ label orbital degree of freedom, $\textit{s}_{3} = 1 (-1)$ represents spin-up (down) component, and $r_s=2-s_3$ stands for the radius of nodal lines. We first put forward a general framework to classify two kinds of NLSGSMs (Fig.~\ref{fig1}), and tabulate the corresponding effective Hamiltonians and the connection modes of NLSGSMs in Table~\ref{Table1}. Regarding type-A NLSGSMs (Fig.~\ref{fig1}(a)), opposite spin-polarized states coexist at the Fermi level, which greatly decreases the pure spin-polarized current. Moreover, the hybrid spin current would restrict the further practicability of NLSGSMs. By comparison, type-B NLSGSMs, as pictured in Fig.~\ref{fig1}(b), which can effectively avoid the shortcoming of the coexistence of the two spin channel currents, which is the focus of our work. 

Compared to Weyl or Dirac semimetal with monotonous structural variations, TNL family can exhibit numerous cross-connection configurations in momentum space, including nodal lines,~\cite{4mullen2015line, 5fang2015topological, 6weng2015topological, 7kim2015dirac, 8yu2015topological, 9ezawa2016loop, 10bian2016topological,11hu2016evidence, 12schoop2016dirac, 13li2016dirac, 14kobayashi2017crossing, 15sun2017dirac, 16li2017type, 17wang2017antiferromagnetic, 18zhang2018nodal, 19ma2018mirror,20gong2018symmorphic} nodal chains,~\cite{21bzduvsek2016nodal, 22wang2017hourglass} nodal links,~\cite{23yan2017nodal, 24chen2017topological} nodal nets~\cite{25fu2018hourglasslike, 26singh2018topological, 27feng2018topological, 28wang2018topological} and nodal knots,~\cite{29ezawa2017topological, 30bi2017nodal}, which will provide more candidates for high-performance spintronic materials. The drumhead-like surface states as a significant indicator of TNLSs, may generate exotic behaviors~\cite{13li2016dirac, 46rhim2015landau, 47huh2016long}. It is desirable to search for large drumhead-like surface states in SGSMs. Via different cross-connection modes, multiple NLSGSMs (\textit{e}.\textit{g}., nodal chain SGSM), as pictured in Fig.~\ref{fig1}(c), can produce a larger surface density of states compared with a single NL, which may offer an effective avenue to study more interesting effects.   

\begin{table}
  \centering
  \caption{Parameters of effective models in two kinds of NLSGSMs.}

  \label{Table1}
  \begin{tabular}{c|c|c|c}
    \hline
    \hline
    \multirowcell{2}{Classification} & \textit{s}$_3$ = 1 & \textit{s}$_3$ = -1 & \multirowcell{2}{Description} \\
    & (spin up) & (spin down) & \\
    \hline
    Type A & \multicolumn{2}{c|}{\textit{m}=0}& Fig.~\ref{fig1}(a) \\
    \hline	
    Type B & \multicolumn{2}{c|}{\textit{m} $\neq$ 0} & Fig.~\ref{fig1}(b)\\
    \hline
    \hline
  \end{tabular}
\end{table}

\begin{figure}
\includegraphics[width=1.0\columnwidth]{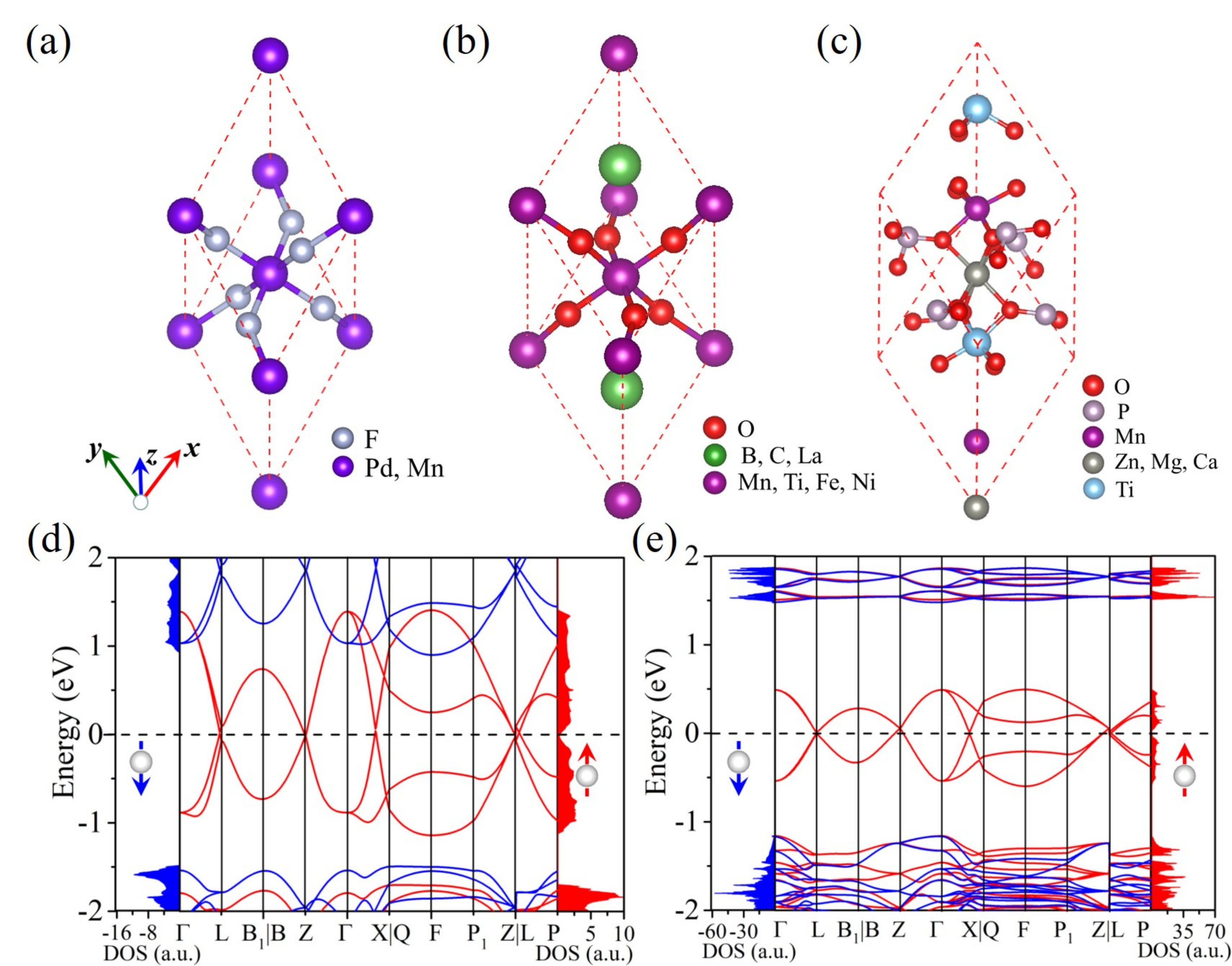}
\caption
{(Color online) (a) The primitive unit cell with nonsymmorphic $\textit{R}\overline{3}\textit{c}$ rhombohedral lattice of TMF$_3$ (TM = Mn, Pd) and (b) $ABO_3$-type perovskites (\textit{i}.\textit{e}., FeCO$_3$, MnBO$_3$, TiBO$_3$, LaMnO$_3$ and LaNiO$_3$). (c) The primitive unit cell with $\textit{R}{3}\textit{c}$ rhombohedral $A^{\prime}ABO_3$-type perovskites (\textit{i}.\textit{e}., XTiMn(PO$_4$)$_3$, X = Ca, Mg, Zn). Spin-resolved band structures and DOS of (d) PdF$_3$ and (e) ZnTiMn(PO$_4$)$_3$  (U=4 eV). Red and blue lines represent spin-up and spin-down channels, respectively.}
\label{fig2}
\end{figure}

\textit{\textcolor{black}{High-quality Candidate Materials.---}} The concrete materials realization plays an important role in the theoretical prototypes into reality. Here we sort out a series of candidates with NLSGSM (chainlike) states that have not been mentioned to date, involving rhombohedral transition metal trifluorides with $\textit{R}\overline{3}\textit{c}$ space group (\textit{i}.\textit{e}., PdF$_3$~\cite{48hepworth1957crystalPdF3} and MnF$_3$~\cite{49muller1987kristallstruktur}) and rhombohedral transition metal carbonate (or borates) with $\textit{R}\overline{3}\textit{c}$ space group (\textit{i}.\textit{e}., FeCO$_3$ (Siderite)~\cite{50effenberger1981crystal}, TiBO$_3$~\cite{51huber1995crystal}, MnBO$_3$, LaMnO$_3$~\cite{52moreno2008preparation} and LaNiO$_3$~\cite{53garcia1992neutron}) as well as rhombohedral transition metal phosphates with $\textit{R}{3}\textit{c}$ space group (\textit{i}.\textit{e}., XTiMn(PO$_4$)$_3$, X = Ca, Mg, Zn), as shown in Fig.~\ref{fig2}(a)-(c), respectively. In these nodal chain SGSMs, they are characterized by ultra-clean energy dispersion and ultra-high Fermi velocity, which will vastly promote the experimental progress of NLSGSMs family. We will focus on two representative candidates from the two distinct space groups, \textit{i}.\textit{e}., PdF$_3$ and ZnTiMn(PO$_4$)$_3$, and further understand the striking physics for the nodal chain SGSMs. Moreover, the remaining candidates with very similar electronic features are provided in~\cite{54}.  

As an ideal nodal chain SGSM candidate, PdF$_3$ has been first identified with powder X-ray diffraction in 1957~\cite{48hepworth1957crystalPdF3} and shares an $\textit{R}\overline{3}\textit{c}$ ($D_{3d}^{6}$, No. 167) space group with a rhombohedral primitive unit cell composed of eight atoms; The optimized lattice constants for PdF$_3$ agree well with the experimental lattice parameters~\cite{48hepworth1957crystalPdF3}, and the computational details are shown in Table S11~\cite{54}. Also, we further screen nodal chain SGSM candidates (\textit{e}.\textit{g}., XTiMn(PO$_4$)$_3$, X = Ca, Mg, Zn) in similar $\textit{R}3\textit{c}$ space group ($C_{3v}^{6}$, No. 161) from the Materials Project~\cite{55jain2013commentary}. These transition metal phosphates pave the way to design novel nodal chain SGSM materials. Regarding the transition metal-based materials, the magnetic property is generally attributed to the transition metal atoms. To determine the magnetic ground states of PdF$_3$ and ZnTiMn(PO$_4$)$_3$, we calculate the total energies of three different magnetic configurations, including FM and two antiferromagnetic (AFM1 and AFM2) configurations, as shown in Fig. S1~\cite{54}. We find that the FM state is lower in energy than AFM1 and AFM2 states, respectively, indicating that the PdF$_3$ and ZnTiMn(PO$_4$)$_3$ prefer FM ground states (see the details in Table S12~\cite{54}). 

Stable FM ground states provide us the motivation to further explore the desired properties for PdF$_3$ and ZnTiMn(PO$_4$)$_3$. In Fig.~\ref{fig2}(d) and Fig.~\ref{fig2}(e), both of the band structures show a linear dispersion semimetallic feature in the spin-up channel, whereas the spin-down channel presents semiconductor with large gaps of 2.39 eV and 2.64 eV. Indeed, from the density of states (DOS) in PdF$_3$ and ZnTiMn(PO$_4$)$_3$, the energy ranges of linear dispersion are enough to overcome the interruption of irrelevant bands, therefore making ideal half-metals with fully spin-polarized. Generally, the bigger slope is expected to be higher Fermi velocity in the linear dispersion system. As for PdF$_3$ and ZnTiMn(PO$_4$)$_3$, the Fermi velocities are estimated about $4.85\times10^{5}$ m/s and $2.98\times10^{5}$ m/s along the $\varGamma\rightarrow X$ path, respectively, which are on the order of graphene, meaning that these candidates would possess more intriguing properties. 

As for $\textit{R}\overline{3}\textit{c}$ space group (\textit{e}.\textit{g}., PdF$_3$), the symmetries include the following important operations: the space-inversion symmetry $\mathcal{P}$, the threefold rotation $\mathcal{C}_{3}$ along the [111] direction, and the glide mirrors ($\mathcal{\widetilde{M}}_{x-y}$, $\mathcal{\widetilde{M}}_{y-z}$ $\mathcal{\widetilde{M}}_{-xz}$). In ZnTiMn(PO$_4$)$_3$, the Ti and Mn sites are inequivalent, leading the $\mathcal{P}$ broken, the $\textit{R}\overline{3}\textit{c}$ transforms into the $\textit{R}{3}\textit{c}$, then the operations reduce to the $\mathcal{C}_{3}$ and the three glide mirrors only. In Fig.~\ref{fig2}(d) and Fig.~\ref{fig2}(e), the crossing-node $Z$ appears at both the PdF$_3$ and the ZnTiMn(PO$_4$)$_3$, which is a noticeable feature in the linear dispersion band structures. To trace this trait, our further analysis shows that the two-dimensional irreducible real representation of $\mathcal{C}_{3}$ symmetry based on $d_{yz}$ and $d_{xz}$ generates two pairs of two-fold degeneracy\cite{56fu2011topological}; Moreover, the $Z$ point is the invariant point of the rotation part of three glide mirrors, and the nonsymmorphic symmetry $\mathcal{\widetilde{C}}_{2(x-y)}$ or $\mathcal{\widetilde{M}}_{x-y}$ constrains the Hamiltonian and leads to another pair of two-fold degeneracy at $Z$ point\cite{57young2015dirac}. Remarkably, the former two-fold degeneracy is from the symmorphic symmetry (same atom's $d-$orbitals), and the latter two-fold degeneracy is from the nonsymmorphic symmetry (different atom's $d-$orbitals). Therefore, the neck crossing-point $Z$ features four-fold degeneracy. 

From the orbital projection analysis, as plotted in Fig.S2~\cite{54}, the low-energy dispersion of the spin-up channel near the Fermi level is mainly dominated by the 4$d$ (3$d$) orbitals of Pd (Mn) atoms. PdF$_3$ and ZnTiMn(PO$_4$)$_3$ manifest the triangular twisted oxygen octahedral crystal feature, which induces Pd (Mn) $d$ orbitals to split into two groups: $A_{1g}$ ($d_{z^2}$) and $E_g$ $\left\{ (d_{x^{2}-y^{2}},d_{xy}),(d_{xz},d_{yz})\right\}$ in the $\textit{R}\overline{3}\textit{c}$ and the $\textit{R}{3}\textit{c}$ space groups. Considering the correlation effects for transition metal elements in PdF$_3$ and ZnTiMn(PO$_4$)$_3$, wide range Hubbard U values ($3\sim5$ eV) are checked in Fig.S3~\cite{54}. We can see that the obvious character of the SGSMs are robust against the various U values.

\begin{figure}
\includegraphics[width=1.0\columnwidth]{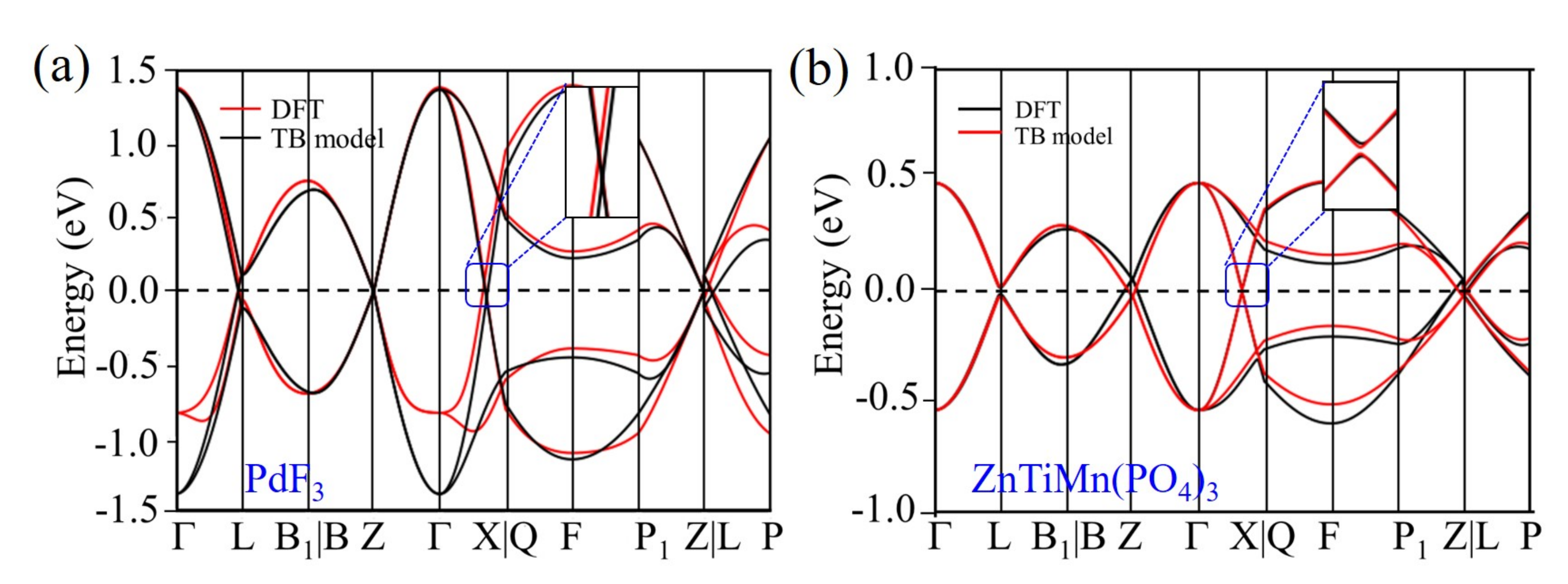}
\caption
{(Color online) The band structures of DFT (black
solid curves) agree with that of the effective lattice model
(red solid curves) calculations in (a) PdF$_3$ and (b) ZnTiMn(PO$_4$)$_3$ systems.}
\label{fig3}
\end{figure}

In the spin-polarized system, spin is a good quantum number, and degrees of freedoms of the spin and the orbital are independent, therefore the crystalline symmetries for the single spin channel can be preserved~\cite{58wang2016time}. To better capture the key physics underlying nodal chain SGSM, we develop a TB model by considering \textit{d}-\textit{d} hoppings using the minimal set of $d_{xz}$ and $d_{yz}$ orbitals as bases. To conveniently describe the atomic bases via the $\textit{R}\overline{3}\textit{c}$ and the $\textit{R}{3}\textit{c}$ space groups, we denote them as $\varphi_1=\ket{d_{xz}},\varphi_2=\ket{d_{yz}}$, and consider the nearest-neighbor(NN) and the next-nearest-neighbor (NNN) hoppings, the TB model Hamiltonian can be expressed as
\begin{equation}
\begin{aligned}
H_{lml'm'}(\boldsymbol{k})=&\sum_{\boldsymbol{d}_j}e^{i\boldsymbol{k \cdot d}_j}E_{lml'm'}(\boldsymbol{d}_j)\\
E_{lm,l'm'}(\boldsymbol{d}_j)=&\braket{\varphi_{m}(\boldsymbol{r}-\boldsymbol{d}_l)|H|\varphi_{m'}(\boldsymbol{r}-\boldsymbol{d}_{l'}-\boldsymbol{R}_j)}, \label{Hami}
\end{aligned}
\end{equation}
where $l$ is the atom index, $E_{lml'm'}(\boldsymbol{d}_j)$ denotes the hopping integrals for neighboring sites with displacement $\boldsymbol{d}_j$. The hopping
integrals to $R\boldsymbol{d}_{\tilde{j}}$ sites can be generated by $\boldsymbol{d}_{j}$~\cite{59gresch2018automated, 60liu2013three}
\begin{equation}
\label{import}
E_{\tilde{l}m\tilde{l'}m'}(R\boldsymbol{d}_{\tilde{j}})=D(R)E_{lml'm'}(\boldsymbol{d}_j)[D(R)]^{\dag},
\end{equation}
where $R$ is the rotation part of symmetry operator, $\boldsymbol{d}_{\tilde{l}}=\{R|\boldsymbol{t}\}\boldsymbol{d}_{l}$, and $D(R)$ is the representation matrix of the $E_g$ irreducible representation. The detailed Hamiltonians of PdF$_3$ and ZnTiMn(PO$_4$)$_3$ are given in Model Section~\cite{54}. The band structures calculated by the TB model Hamiltonian agree well with the ones obtained by the DFT method in the whole BZ, as shown in Fig.~\ref{fig3}(a) and Fig~\ref{fig3}(b). Interestingly, the spin-resolved band structures of PdF$_3$ (Fig.~\ref{fig3}(a)) and ZnTiMn(PO$_4$)$_3$ (Fig.~\ref{fig3}(b)) are similar, except that ZnTiMn(PO$_4$)$_3$ has a tiny gap along the $\varGamma\rightarrow X$ path due to the $\mathcal{P}$ is broken. 

\begin{figure}
\includegraphics[width=1.0\columnwidth]{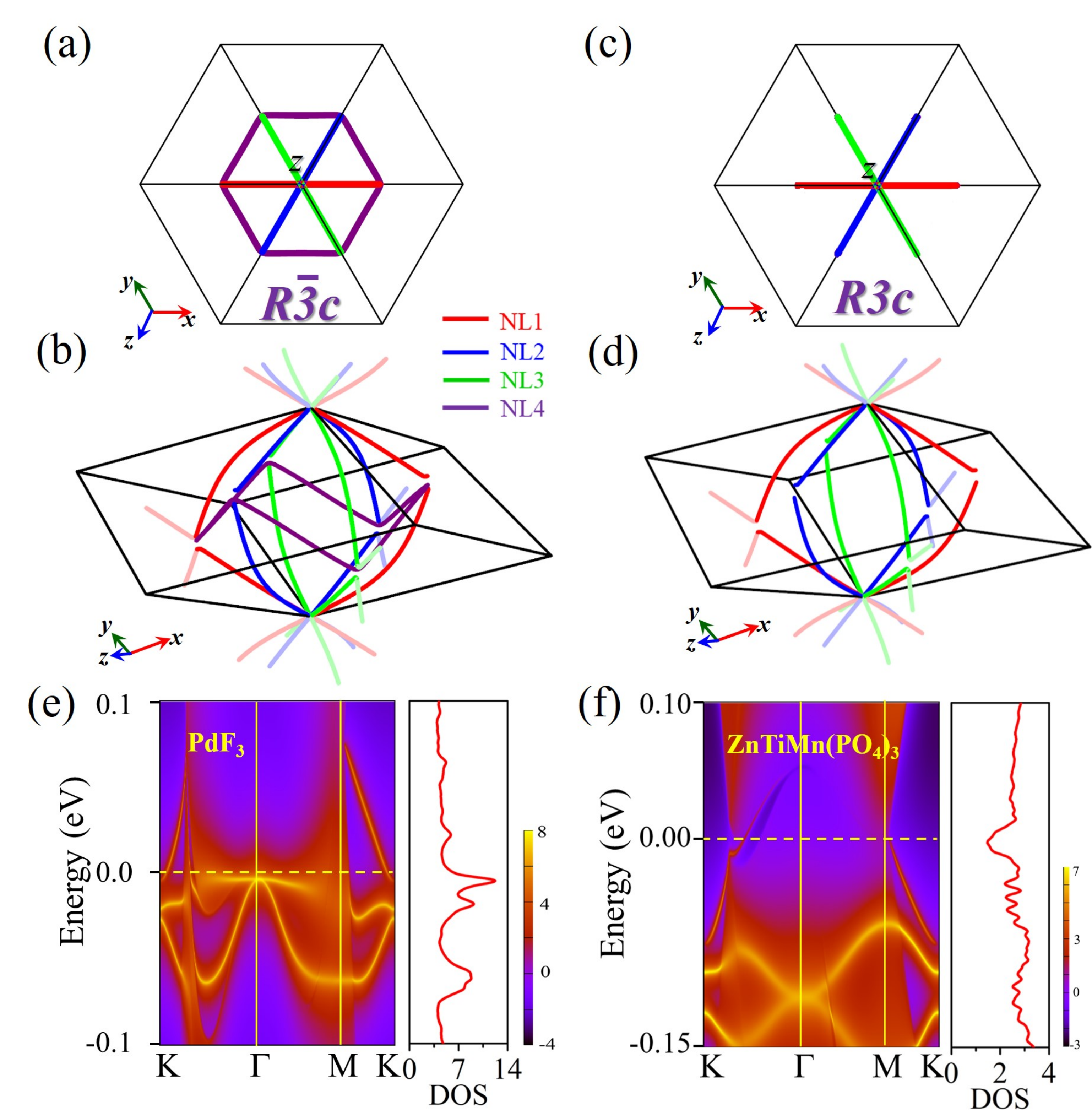}
\caption
{(Color online) (a-d) The profile of the nodal chainlike SGSMs in three-dimensional \textit{k} space, including the top views (a) PdF$_3$ and (c) ZnTiMn(PO$_4$)$_3$ along the [111] direction, the side views (b) PdF$_3$ and (d) ZnTiMn(PO$_4$)$_3$. Diﬀerent colors (\textit{i}.\textit{e}., red line, blue line, green line and purple line) correspond to different orientations of the NLs (\textit{i}.\textit{e}., NL1, NL2, NL3 and NL4). Band structures and DOS for surface(111) (e) and (f) in PdF$_3$ and ZnTiMn(PO$_4$)$_3$. The purple dashed lines labeled \textit{L}$_1$, \textit{L}$_2$ and \textit{L}$_3$ represent three respective loops along which the Berry phase is calculated.}
\label{fig4}
\end{figure}

The evolution of crossing nodes develops two kinds of cross-connection modes under the $\textit{R}\overline{3}\textit{c}$ constraint, \textit{i}.\textit{e}., nodal chains and NLs. In PdF$_3$, the profile of all nodes is revealed by TB model clearly, as shown in Fig.~\ref{fig4}(a) and Fig~\ref{fig4}(b). For Case-I, the cross-connection structure (see Fig.~\ref{fig4}(b)) with three NLs (\textit{i}.\textit{e}., NL1, NL2, and NL3) are pinned at the $Z$ point to form a nodal chainlike structure. The neck crossing-point $Z$ is co-constrained by the $\mathcal{C}_{3}$ rotation symmetry along $\Gamma-Z$ high symmetry line and the nonsymmorphic symmetry $\mathcal{\widetilde{C}}_{2(x-y)}$ or $\mathcal{\widetilde{M}}_{x-y}$. Remarkably, the NL1 on the glide mirror $\mathcal{\widetilde{M}}_{y-z}$ plane is just accidentally formed rather than protected by the nonsymmorphic symmetry (see Fig. S4 \cite{54}). Combining with $\mathcal{\widetilde{M}}_{y-z}$ and $\mathcal{C}_{3}$ rotation symmetries, the NL2 and NL3 possess the same characteristics. Besides, as seen in Fig.~\ref{fig4}(b), akin to alkaline-earth compounds~\cite{61huang2016topological}, the case-II cross-connection structure (\textit{i}.\textit{e}., NL4) is protected by $\mathcal{P}$ and the complex conjugate operator $\mathcal{K}$. While, the ``snake"-like NL4 can be annihilated in ZnTiMn(PO$_4$)$_3$ since the $\mathcal{P}$ disappears, as displayed in Fig.~\ref{fig4}(c) and Fig.~\ref{fig4}(d). The topological protection of NL1 (NL2 or NL3 or NL4) are further checked by directly calculating the nontrivial $\pi$ Berry phase along a small loop (\textit{L}$_1$) enclosing the NL. Moreover, based on \textit{L}$_2$ (or \textit{L}$_3$) encircling the number of the nodal chain (two NLs or three NLs), zero (or $\pi$) Berry phase is obtained. As a result, the origin of multiple lines (nodal chain and NL) is revealed and the detailed analysis is explained in the Model Section~\cite{54}. 

\textit{\textcolor{black}{Discussion.---}}We further investigated the robustness and the feasibility of the nodal chainlike SGSMs. Taking PdF$_3$ as an example, we impose a small perturbation (\emph{e}.\emph{g}. external triaxial strain) on it. Compared with the fragile NLSs, the band structures of PdF$_3$ under such a wide range of triaxial compressional strains ($0\sim-10$\%), all crossing nodes still exist as expected, manifesting the robustness of Weyl nodal chain against certain structural deformation (see the details in Fig. S5~\cite{54}). 

Practically, as an essential property of magnetic material PdF$_3$, magnetic anisotropy has a great significance for aligning the magnetic moments in magnetic storage media. To further determine the magnetocrystalline anisotropy energy (MAE) of PdF$_3$, we calculate the angular dependence of the MAE on the magnetization angle $\theta$ as a function of polar angles in different directions on the \emph{x}(\emph{y})-\emph{z} and \emph{x}-\emph{y} planes, as plotted in Fig. S6~\cite{54}. It is obvious that MAE is nearly equivalent to the $\phi$ evolution in the \emph{x}-\emph{y} plane. Regarding the \emph{x}(\emph{y})-\emph{z}, the MAE of PdF$_3$ can reach the maximal value 1.19 meV/atom at the $\theta=0$ (degree), which is one order of magnitude larger than that the MAE in cubic Fe, Co, and Ni~\cite{62halilov1998magnetocrystalline} meaning PdF$_3$ prefers FM state with the spin-aligned along the \emph{z}-direction in the conventional cell (\textit{i}.\textit{e}., the [111] in the unit cell).  

The hallmark drumhead surface states for topological NLSGSMs PdF$_3$ and ZnTiMn(PO$_4$)$_3$  are shown in Fig.~\ref{fig4}(e) and Fig.~\ref{fig4}(f). Due to the lack of chiral symmetry or particle-hole symmetry in real solid materials, the drumhead states are dispersive indeed and the dispersion strength is material dependent. The peaklike DOS from the nearly flat drumhead states are clearly shown for PdF$_3$, as plotted in Fig.~\ref{fig4}(e). Volovik \textit{et} \textit{al}. proposed that 2D flat bands with large DOS provide a route to achieving high-temperature superconductivity\cite{63kopnin2011high, 64volovik2015standard, 65heikkila2016flat}. Besides, recent experimental and theoretical advances\cite{58wang2016time, 66chang2016room, 67chang2017topological, 68xu2018topological, 69noky2019large,70belopolski2019discovery, 71liu2019magnetic, 72Morali1286} in the nodal half-metal states oﬀered a tremendous boost to the emerging field of the magnetic semimetal's community. Herein, we would like to stress that the 2D nearly flat drumhead states in TNLSGSMs not only have large DOS but also are fully spin-polarized. Such fully spin-polarized nearly flat drumhead states would lead to the equal spin pairing topological superconductivity, such as $p$ or $f$ wave topological superconductivity, which supports Majorana zero bound states at vortices with non-Abelian statistics for the intriguing proposal of topological quantum computation. Recently, topological catalysts provide a potential platform to create active sites\cite{73chen2011co, 74rajamathi2017weyl,75li2018topological}. PdF$_3$ as a Pd-based noble metal binary compound, which would be a better electrocatalyst for the hydrogen evolution reaction (HER) due to following distinguishing aspects: (i) Pd itself is a good catalyst; (ii) topological nodal chain induced large drumhead surface states can offer sufficient active planes; (iii) SGSM featured linear crossings of energy bands can provide high carrier mobility near Fermi level. Therefore, PdF$_3$ showcases a new routine to design a promising electrocatalyst. 

\textit{\textcolor{black}{Conclusion.---}}We first introduce a general framework to classify opposite and same spin-polarized Weyl nodal chainlike states in SGSMs, and propose a series of realistic materials to realize the hitherto unreported the fully spin-polarized nodal chain states. These high-quality materials harbor ultra-clean energy dispersion and ultra-high Fermi velocity, which is rather robust against strong triaxial compressional strain. The proposal of TNLSGSMs and their materials realization greatly expand TSMs family and provide a good playground for spintronics, topological superconductivity, and topological catalysts. 
%

%~\nocite{bzduvsek2016nodal, zhang2018ideal, heikkila2011flat, perdew1996generalized, blochl1994projector, peng2016versatile, 69, 70, 71, 72}%

The work is supported by the National Natural Science Foundation of China (Grants Nos. 11774028, 11734003, 11574029), the National Key R\&D Program of China (Grant No. 2016YFA0300600), the Strategic Priority Research Program of Chinese Academy of Sciences (Grant No. XDB30000000), and Basic Research Funds of Beijing Institute of Technology (No. 2017CX01018). R.W.Z. also thanks the supports from Graduate Technological Innovation Project of Beijing Institute of Technology (Grant No. 2018CX10028, 2019CX10018).
\bibliography{ref}

\begin{thebibliography}{75}%
\makeatletter
\providecommand \@ifxundefined [1]{%
 \@ifx{#1\undefined}
}%
\providecommand \@ifnum [1]{%
 \ifnum #1\expandafter \@firstoftwo
 \else \expandafter \@secondoftwo
 \fi
}%
\providecommand \@ifx [1]{%
 \ifx #1\expandafter \@firstoftwo
 \else \expandafter \@secondoftwo
 \fi
}%
\providecommand \natexlab [1]{#1}%
\providecommand \enquote  [1]{``#1''}%
\providecommand \bibnamefont  [1]{#1}%
\providecommand \bibfnamefont [1]{#1}%
\providecommand \citenamefont [1]{#1}%
\providecommand \href@noop [0]{\@secondoftwo}%
\providecommand \href [0]{\begingroup \@sanitize@url \@href}%
\providecommand \@href[1]{\@@startlink{#1}\@@href}%
\providecommand \@@href[1]{\endgroup#1\@@endlink}%
\providecommand \@sanitize@url [0]{\catcode `\\12\catcode `\$12\catcode
  `\&12\catcode `\#12\catcode `\^12\catcode `\_12\catcode `\%12\relax}%
\providecommand \@@startlink[1]{}%
\providecommand \@@endlink[0]{}%
\providecommand \url  [0]{\begingroup\@sanitize@url \@url }%
\providecommand \@url [1]{\endgroup\@href {#1}{\urlprefix }}%
\providecommand \urlprefix  [0]{URL }%
\providecommand \Eprint [0]{\href }%
\providecommand \doibase [0]{http://dx.doi.org/}%
\providecommand \selectlanguage [0]{\@gobble}%
\providecommand \bibinfo  [0]{\@secondoftwo}%
\providecommand \bibfield  [0]{\@secondoftwo}%
\providecommand \translation [1]{[#1]}%
\providecommand \BibitemOpen [0]{}%
\providecommand \bibitemStop [0]{}%
\providecommand \bibitemNoStop [0]{.\EOS\space}%
\providecommand \EOS [0]{\spacefactor3000\relax}%
\providecommand \BibitemShut  [1]{\csname bibitem#1\endcsname}%
\let\auto@bib@innerbib\@empty
%</preamble>
\bibitem [{\citenamefont {Wolf}\ \emph {et~al.}(2001)\citenamefont {Wolf},
  \citenamefont {Awschalom}, \citenamefont {Buhrman}, \citenamefont {Daughton},
  \citenamefont {von Moln{\'a}r}, \citenamefont {Roukes}, \citenamefont
  {Chtchelkanova},\ and\ \citenamefont {Treger}}]{1wolf2001spintronics}%
  \BibitemOpen
  \bibfield  {author} {\bibinfo {author} {\bibfnamefont {S.~A.}\ \bibnamefont
  {Wolf}}, \bibinfo {author} {\bibfnamefont {D.~D.}\ \bibnamefont {Awschalom}},
  \bibinfo {author} {\bibfnamefont {R.~A.}\ \bibnamefont {Buhrman}}, \bibinfo
  {author} {\bibfnamefont {J.~M.}\ \bibnamefont {Daughton}}, \bibinfo {author}
  {\bibfnamefont {S.}~\bibnamefont {von Moln{\'a}r}}, \bibinfo {author}
  {\bibfnamefont {M.~L.}\ \bibnamefont {Roukes}}, \bibinfo {author}
  {\bibfnamefont {A.~Y.}\ \bibnamefont {Chtchelkanova}}, \ and\ \bibinfo
  {author} {\bibfnamefont {D.~M.}\ \bibnamefont {Treger}},\ }\href@noop {}
  {\bibfield  {journal} {\bibinfo  {journal} {Science}\ }\textbf {\bibinfo
  {volume} {294}},\ \bibinfo {pages} {1488} (\bibinfo {year}
  {2001})}\BibitemShut {NoStop}%
\bibitem [{\citenamefont {\ifmmode \check{Z}\else
  \v{Z}\fi{}uti\ifmmode~\acute{c}\else \'{c}\fi{}}\ \emph
  {et~al.}(2004)\citenamefont {\ifmmode \check{Z}\else
  \v{Z}\fi{}uti\ifmmode~\acute{c}\else \'{c}\fi{}}, \citenamefont {Fabian},\
  and\ \citenamefont {Das~Sarma}}]{2vzutic2004spintronics}%
  \BibitemOpen
  \bibfield  {author} {\bibinfo {author} {\bibfnamefont {I.}~\bibnamefont
  {\ifmmode \check{Z}\else \v{Z}\fi{}uti\ifmmode~\acute{c}\else \'{c}\fi{}}},
  \bibinfo {author} {\bibfnamefont {J.}~\bibnamefont {Fabian}}, \ and\ \bibinfo
  {author} {\bibfnamefont {S.}~\bibnamefont {Das~Sarma}},\ }\href@noop {}
  {\bibfield  {journal} {\bibinfo  {journal} {Rev. Mod. Phys.}\ }\textbf
  {\bibinfo {volume} {76}},\ \bibinfo {pages} {323} (\bibinfo {year}
  {2004})}\BibitemShut {NoStop}%
\bibitem [{\citenamefont {Wang}(2008)}]{3wang2008proposal}%
  \BibitemOpen
  \bibfield  {author} {\bibinfo {author} {\bibfnamefont {X.}~\bibnamefont
  {Wang}},\ }\href@noop {} {\bibfield  {journal} {\bibinfo  {journal} {Phys.
  Rev. Lett.}\ }\textbf {\bibinfo {volume} {100}},\ \bibinfo {pages} {156404}
  (\bibinfo {year} {2008})}\BibitemShut {NoStop}%
\bibitem [{\citenamefont {Mullen}\ \emph {et~al.}(2015)\citenamefont {Mullen},
  \citenamefont {Uchoa},\ and\ \citenamefont {Glatzhofer}}]{4mullen2015line}%
  \BibitemOpen
  \bibfield  {author} {\bibinfo {author} {\bibfnamefont {K.}~\bibnamefont
  {Mullen}}, \bibinfo {author} {\bibfnamefont {B.}~\bibnamefont {Uchoa}}, \
  and\ \bibinfo {author} {\bibfnamefont {D.~T.}\ \bibnamefont {Glatzhofer}},\
  }\href@noop {} {\bibfield  {journal} {\bibinfo  {journal} {Phys. Rev. Lett.}\
  }\textbf {\bibinfo {volume} {115}},\ \bibinfo {pages} {026403} (\bibinfo
  {year} {2015})}\BibitemShut {NoStop}%
\bibitem [{\citenamefont {Fang}\ \emph {et~al.}(2015)\citenamefont {Fang},
  \citenamefont {Chen}, \citenamefont {Kee},\ and\ \citenamefont
  {Fu}}]{5fang2015topological}%
  \BibitemOpen
  \bibfield  {author} {\bibinfo {author} {\bibfnamefont {C.}~\bibnamefont
  {Fang}}, \bibinfo {author} {\bibfnamefont {Y.}~\bibnamefont {Chen}}, \bibinfo
  {author} {\bibfnamefont {H.-Y.}\ \bibnamefont {Kee}}, \ and\ \bibinfo
  {author} {\bibfnamefont {L.}~\bibnamefont {Fu}},\ }\href@noop {} {\bibfield
  {journal} {\bibinfo  {journal} {Phys. Rev. B}\ }\textbf {\bibinfo {volume}
  {92}},\ \bibinfo {pages} {081201} (\bibinfo {year} {2015})}\BibitemShut
  {NoStop}%
\bibitem [{\citenamefont {Weng}\ \emph
  {et~al.}(2015{\natexlab{a}})\citenamefont {Weng}, \citenamefont {Liang},
  \citenamefont {Xu}, \citenamefont {Yu}, \citenamefont {Fang}, \citenamefont
  {Dai},\ and\ \citenamefont {Kawazoe}}]{6weng2015topological}%
  \BibitemOpen
  \bibfield  {author} {\bibinfo {author} {\bibfnamefont {H.}~\bibnamefont
  {Weng}}, \bibinfo {author} {\bibfnamefont {Y.}~\bibnamefont {Liang}},
  \bibinfo {author} {\bibfnamefont {Q.}~\bibnamefont {Xu}}, \bibinfo {author}
  {\bibfnamefont {R.}~\bibnamefont {Yu}}, \bibinfo {author} {\bibfnamefont
  {Z.}~\bibnamefont {Fang}}, \bibinfo {author} {\bibfnamefont {X.}~\bibnamefont
  {Dai}}, \ and\ \bibinfo {author} {\bibfnamefont {Y.}~\bibnamefont
  {Kawazoe}},\ }\href@noop {} {\bibfield  {journal} {\bibinfo  {journal} {Phys.
  Rev. B}\ }\textbf {\bibinfo {volume} {92}},\ \bibinfo {pages} {045108}
  (\bibinfo {year} {2015}{\natexlab{a}})}\BibitemShut {NoStop}%
\bibitem [{\citenamefont {Kim}\ \emph {et~al.}(2015)\citenamefont {Kim},
  \citenamefont {Wieder}, \citenamefont {Kane},\ and\ \citenamefont
  {Rappe}}]{7kim2015dirac}%
  \BibitemOpen
  \bibfield  {author} {\bibinfo {author} {\bibfnamefont {Y.}~\bibnamefont
  {Kim}}, \bibinfo {author} {\bibfnamefont {B.~J.}\ \bibnamefont {Wieder}},
  \bibinfo {author} {\bibfnamefont {C.~L.}\ \bibnamefont {Kane}}, \ and\
  \bibinfo {author} {\bibfnamefont {A.~M.}\ \bibnamefont {Rappe}},\ }\href@noop
  {} {\bibfield  {journal} {\bibinfo  {journal} {Phys. Rev. Lett.}\ }\textbf
  {\bibinfo {volume} {115}},\ \bibinfo {pages} {036806} (\bibinfo {year}
  {2015})}\BibitemShut {NoStop}%
\bibitem [{\citenamefont {Yu}\ \emph {et~al.}(2015)\citenamefont {Yu},
  \citenamefont {Weng}, \citenamefont {Fang}, \citenamefont {Dai},\ and\
  \citenamefont {Hu}}]{8yu2015topological}%
  \BibitemOpen
  \bibfield  {author} {\bibinfo {author} {\bibfnamefont {R.}~\bibnamefont
  {Yu}}, \bibinfo {author} {\bibfnamefont {H.}~\bibnamefont {Weng}}, \bibinfo
  {author} {\bibfnamefont {Z.}~\bibnamefont {Fang}}, \bibinfo {author}
  {\bibfnamefont {X.}~\bibnamefont {Dai}}, \ and\ \bibinfo {author}
  {\bibfnamefont {X.}~\bibnamefont {Hu}},\ }\href@noop {} {\bibfield  {journal}
  {\bibinfo  {journal} {Phys. Rev. Lett.}\ }\textbf {\bibinfo {volume} {115}},\
  \bibinfo {pages} {036807} (\bibinfo {year} {2015})}\BibitemShut {NoStop}%
\bibitem [{\citenamefont {Ezawa}(2016)}]{9ezawa2016loop}%
  \BibitemOpen
  \bibfield  {author} {\bibinfo {author} {\bibfnamefont {M.}~\bibnamefont
  {Ezawa}},\ }\href@noop {} {\bibfield  {journal} {\bibinfo  {journal} {Phys.
  Rev. Lett.}\ }\textbf {\bibinfo {volume} {116}},\ \bibinfo {pages} {127202}
  (\bibinfo {year} {2016})}\BibitemShut {NoStop}%
\bibitem [{\citenamefont {Bian}\ \emph {et~al.}(2016)\citenamefont {Bian},
  \citenamefont {Chang}, \citenamefont {Sankar}, \citenamefont {Xu},
  \citenamefont {Zheng}, \citenamefont {Neupert}, \citenamefont {Chiu},
  \citenamefont {Huang}, \citenamefont {Chang}, \citenamefont {Belopolski}
  \emph {et~al.}}]{10bian2016topological}%
  \BibitemOpen
  \bibfield  {author} {\bibinfo {author} {\bibfnamefont {G.}~\bibnamefont
  {Bian}}, \bibinfo {author} {\bibfnamefont {T.-R.}\ \bibnamefont {Chang}},
  \bibinfo {author} {\bibfnamefont {R.}~\bibnamefont {Sankar}}, \bibinfo
  {author} {\bibfnamefont {S.-Y.}\ \bibnamefont {Xu}}, \bibinfo {author}
  {\bibfnamefont {H.}~\bibnamefont {Zheng}}, \bibinfo {author} {\bibfnamefont
  {T.}~\bibnamefont {Neupert}}, \bibinfo {author} {\bibfnamefont {C.-K.}\
  \bibnamefont {Chiu}}, \bibinfo {author} {\bibfnamefont {S.-M.}\ \bibnamefont
  {Huang}}, \bibinfo {author} {\bibfnamefont {G.}~\bibnamefont {Chang}},
  \bibinfo {author} {\bibfnamefont {I.}~\bibnamefont {Belopolski}},  \emph
  {et~al.},\ }\href@noop {} {\bibfield  {journal} {\bibinfo  {journal} {Nat.
  Commun.}\ }\textbf {\bibinfo {volume} {7}},\ \bibinfo {pages} {10556}
  (\bibinfo {year} {2016})}\BibitemShut {NoStop}%
\bibitem [{\citenamefont {Hu}\ \emph {et~al.}(2016)\citenamefont {Hu},
  \citenamefont {Tang}, \citenamefont {Liu}, \citenamefont {Liu}, \citenamefont
  {Zhu}, \citenamefont {Graf}, \citenamefont {Myhro}, \citenamefont {Tran},
  \citenamefont {Lau}, \citenamefont {Wei} \emph {et~al.}}]{11hu2016evidence}%
  \BibitemOpen
  \bibfield  {author} {\bibinfo {author} {\bibfnamefont {J.}~\bibnamefont
  {Hu}}, \bibinfo {author} {\bibfnamefont {Z.}~\bibnamefont {Tang}}, \bibinfo
  {author} {\bibfnamefont {J.}~\bibnamefont {Liu}}, \bibinfo {author}
  {\bibfnamefont {X.}~\bibnamefont {Liu}}, \bibinfo {author} {\bibfnamefont
  {Y.}~\bibnamefont {Zhu}}, \bibinfo {author} {\bibfnamefont {D.}~\bibnamefont
  {Graf}}, \bibinfo {author} {\bibfnamefont {K.}~\bibnamefont {Myhro}},
  \bibinfo {author} {\bibfnamefont {S.}~\bibnamefont {Tran}}, \bibinfo {author}
  {\bibfnamefont {C.~N.}\ \bibnamefont {Lau}}, \bibinfo {author} {\bibfnamefont
  {J.}~\bibnamefont {Wei}},  \emph {et~al.},\ }\href@noop {} {\bibfield
  {journal} {\bibinfo  {journal} {Phys. Rev. Lett.}\ }\textbf {\bibinfo
  {volume} {117}},\ \bibinfo {pages} {016602} (\bibinfo {year}
  {2016})}\BibitemShut {NoStop}%
\bibitem [{\citenamefont {Schoop}\ \emph {et~al.}(2016)\citenamefont {Schoop},
  \citenamefont {Ali}, \citenamefont {Straßer}, \citenamefont {Topp},
  \citenamefont {Varykhalov}, \citenamefont {Marchenko}, \citenamefont
  {Duppel}, \citenamefont {Parkin}, \citenamefont {Lotsch},\ and\ \citenamefont
  {Ast}}]{12schoop2016dirac}%
  \BibitemOpen
  \bibfield  {author} {\bibinfo {author} {\bibfnamefont {L.~M.}\ \bibnamefont
  {Schoop}}, \bibinfo {author} {\bibfnamefont {M.~N.}\ \bibnamefont {Ali}},
  \bibinfo {author} {\bibfnamefont {C.}~\bibnamefont {Straßer}}, \bibinfo
  {author} {\bibfnamefont {A.}~\bibnamefont {Topp}}, \bibinfo {author}
  {\bibfnamefont {A.}~\bibnamefont {Varykhalov}}, \bibinfo {author}
  {\bibfnamefont {D.}~\bibnamefont {Marchenko}}, \bibinfo {author}
  {\bibfnamefont {V.}~\bibnamefont {Duppel}}, \bibinfo {author} {\bibfnamefont
  {S.~S.~P.}\ \bibnamefont {Parkin}}, \bibinfo {author} {\bibfnamefont {B.~V.}\
  \bibnamefont {Lotsch}}, \ and\ \bibinfo {author} {\bibfnamefont {C.~R.}\
  \bibnamefont {Ast}},\ }\href@noop {} {\bibfield  {journal} {\bibinfo
  {journal} {Nat. Commun.}\ }\textbf {\bibinfo {volume} {7}},\ \bibinfo {pages}
  {11696} (\bibinfo {year} {2016})}\BibitemShut {NoStop}%
\bibitem [{\citenamefont {Li}\ \emph {et~al.}(2016)\citenamefont {Li},
  \citenamefont {Ma}, \citenamefont {Cheng}, \citenamefont {Wang},
  \citenamefont {Li}, \citenamefont {Zhang}, \citenamefont {Li},\ and\
  \citenamefont {Chen}}]{13li2016dirac}%
  \BibitemOpen
  \bibfield  {author} {\bibinfo {author} {\bibfnamefont {R.}~\bibnamefont
  {Li}}, \bibinfo {author} {\bibfnamefont {H.}~\bibnamefont {Ma}}, \bibinfo
  {author} {\bibfnamefont {X.}~\bibnamefont {Cheng}}, \bibinfo {author}
  {\bibfnamefont {S.}~\bibnamefont {Wang}}, \bibinfo {author} {\bibfnamefont
  {D.}~\bibnamefont {Li}}, \bibinfo {author} {\bibfnamefont {Z.}~\bibnamefont
  {Zhang}}, \bibinfo {author} {\bibfnamefont {Y.}~\bibnamefont {Li}}, \ and\
  \bibinfo {author} {\bibfnamefont {X.-Q.}\ \bibnamefont {Chen}},\ }\href@noop
  {} {\bibfield  {journal} {\bibinfo  {journal} {Phys. Rev. Lett.}\ }\textbf
  {\bibinfo {volume} {117}},\ \bibinfo {pages} {096401} (\bibinfo {year}
  {2016})}\BibitemShut {NoStop}%
\bibitem [{\citenamefont {Kobayashi}\ \emph {et~al.}(2017)\citenamefont
  {Kobayashi}, \citenamefont {Yamakawa}, \citenamefont {Yamakage},
  \citenamefont {Inohara}, \citenamefont {Okamoto},\ and\ \citenamefont
  {Tanaka}}]{14kobayashi2017crossing}%
  \BibitemOpen
  \bibfield  {author} {\bibinfo {author} {\bibfnamefont {S.}~\bibnamefont
  {Kobayashi}}, \bibinfo {author} {\bibfnamefont {Y.}~\bibnamefont {Yamakawa}},
  \bibinfo {author} {\bibfnamefont {A.}~\bibnamefont {Yamakage}}, \bibinfo
  {author} {\bibfnamefont {T.}~\bibnamefont {Inohara}}, \bibinfo {author}
  {\bibfnamefont {Y.}~\bibnamefont {Okamoto}}, \ and\ \bibinfo {author}
  {\bibfnamefont {Y.}~\bibnamefont {Tanaka}},\ }\href@noop {} {\bibfield
  {journal} {\bibinfo  {journal} {Phys. Rev. B}\ }\textbf {\bibinfo {volume}
  {95}},\ \bibinfo {pages} {245208} (\bibinfo {year} {2017})}\BibitemShut
  {NoStop}%
\bibitem [{\citenamefont {Sun}\ \emph {et~al.}(2017)\citenamefont {Sun},
  \citenamefont {Zhang}, \citenamefont {Liu}, \citenamefont {Felser},\ and\
  \citenamefont {Yan}}]{15sun2017dirac}%
  \BibitemOpen
  \bibfield  {author} {\bibinfo {author} {\bibfnamefont {Y.}~\bibnamefont
  {Sun}}, \bibinfo {author} {\bibfnamefont {Y.}~\bibnamefont {Zhang}}, \bibinfo
  {author} {\bibfnamefont {C.-X.}\ \bibnamefont {Liu}}, \bibinfo {author}
  {\bibfnamefont {C.}~\bibnamefont {Felser}}, \ and\ \bibinfo {author}
  {\bibfnamefont {B.}~\bibnamefont {Yan}},\ }\href@noop {} {\bibfield
  {journal} {\bibinfo  {journal} {Phys. Rev. B}\ }\textbf {\bibinfo {volume}
  {95}},\ \bibinfo {pages} {235104} (\bibinfo {year} {2017})}\BibitemShut
  {NoStop}%
\bibitem [{\citenamefont {Li}\ \emph {et~al.}(2017)\citenamefont {Li},
  \citenamefont {Yu}, \citenamefont {Liu}, \citenamefont {Guan}, \citenamefont
  {Wang}, \citenamefont {Zhang}, \citenamefont {Yao},\ and\ \citenamefont
  {Yang}}]{16li2017type}%
  \BibitemOpen
  \bibfield  {author} {\bibinfo {author} {\bibfnamefont {S.}~\bibnamefont
  {Li}}, \bibinfo {author} {\bibfnamefont {Z.-M.}\ \bibnamefont {Yu}}, \bibinfo
  {author} {\bibfnamefont {Y.}~\bibnamefont {Liu}}, \bibinfo {author}
  {\bibfnamefont {S.}~\bibnamefont {Guan}}, \bibinfo {author} {\bibfnamefont
  {S.-S.}\ \bibnamefont {Wang}}, \bibinfo {author} {\bibfnamefont
  {X.}~\bibnamefont {Zhang}}, \bibinfo {author} {\bibfnamefont
  {Y.}~\bibnamefont {Yao}}, \ and\ \bibinfo {author} {\bibfnamefont {S.~A.}\
  \bibnamefont {Yang}},\ }\href@noop {} {\bibfield  {journal} {\bibinfo
  {journal} {Phys. Rev. B}\ }\textbf {\bibinfo {volume} {96}},\ \bibinfo
  {pages} {081106} (\bibinfo {year} {2017})}\BibitemShut {NoStop}%
\bibitem [{\citenamefont {Wang}(2017)}]{17wang2017antiferromagnetic}%
  \BibitemOpen
  \bibfield  {author} {\bibinfo {author} {\bibfnamefont {J.}~\bibnamefont
  {Wang}},\ }\href@noop {} {\bibfield  {journal} {\bibinfo  {journal} {Phys.
  Rev. B}\ }\textbf {\bibinfo {volume} {96}},\ \bibinfo {pages} {081107}
  (\bibinfo {year} {2017})}\BibitemShut {NoStop}%
\bibitem [{\citenamefont {Zhang}\ \emph {et~al.}(2018)\citenamefont {Zhang},
  \citenamefont {Liu}, \citenamefont {Ma}, \citenamefont {Wang},\ and\
  \citenamefont {Yao}}]{18zhang2018nodal}%
  \BibitemOpen
  \bibfield  {author} {\bibinfo {author} {\bibfnamefont {R.-W.}\ \bibnamefont
  {Zhang}}, \bibinfo {author} {\bibfnamefont {C.-C.}\ \bibnamefont {Liu}},
  \bibinfo {author} {\bibfnamefont {D.-S.}\ \bibnamefont {Ma}}, \bibinfo
  {author} {\bibfnamefont {M.}~\bibnamefont {Wang}}, \ and\ \bibinfo {author}
  {\bibfnamefont {Y.}~\bibnamefont {Yao}},\ }\href@noop {} {\bibfield
  {journal} {\bibinfo  {journal} {Phys. Rev. B}\ }\textbf {\bibinfo {volume}
  {98}},\ \bibinfo {pages} {035144} (\bibinfo {year} {2018})}\BibitemShut
  {NoStop}%
\bibitem [{\citenamefont {Ma}\ \emph {et~al.}(2018)\citenamefont {Ma},
  \citenamefont {Zhou}, \citenamefont {Fu}, \citenamefont {Yu}, \citenamefont
  {Liu},\ and\ \citenamefont {Yao}}]{19ma2018mirror}%
  \BibitemOpen
  \bibfield  {author} {\bibinfo {author} {\bibfnamefont {D.-S.}\ \bibnamefont
  {Ma}}, \bibinfo {author} {\bibfnamefont {J.}~\bibnamefont {Zhou}}, \bibinfo
  {author} {\bibfnamefont {B.}~\bibnamefont {Fu}}, \bibinfo {author}
  {\bibfnamefont {Z.-M.}\ \bibnamefont {Yu}}, \bibinfo {author} {\bibfnamefont
  {C.-C.}\ \bibnamefont {Liu}}, \ and\ \bibinfo {author} {\bibfnamefont
  {Y.}~\bibnamefont {Yao}},\ }\href@noop {} {\bibfield  {journal} {\bibinfo
  {journal} {Phys. Rev. B}\ }\textbf {\bibinfo {volume} {98}},\ \bibinfo
  {pages} {201104} (\bibinfo {year} {2018})}\BibitemShut {NoStop}%
\bibitem [{\citenamefont {Gong}\ \emph {et~al.}(2018)\citenamefont {Gong},
  \citenamefont {Xie}, \citenamefont {Chen}, \citenamefont {Kim},\ and\
  \citenamefont {Vanderbilt}}]{20gong2018symmorphic}%
  \BibitemOpen
  \bibfield  {author} {\bibinfo {author} {\bibfnamefont {C.}~\bibnamefont
  {Gong}}, \bibinfo {author} {\bibfnamefont {Y.}~\bibnamefont {Xie}}, \bibinfo
  {author} {\bibfnamefont {Y.}~\bibnamefont {Chen}}, \bibinfo {author}
  {\bibfnamefont {H.-S.}\ \bibnamefont {Kim}}, \ and\ \bibinfo {author}
  {\bibfnamefont {D.}~\bibnamefont {Vanderbilt}},\ }\href@noop {} {\bibfield
  {journal} {\bibinfo  {journal} {Phys. Rev. Lett.}\ }\textbf {\bibinfo
  {volume} {120}},\ \bibinfo {pages} {106403} (\bibinfo {year}
  {2018})}\BibitemShut {NoStop}%
\bibitem [{\citenamefont {Bzdusek}\ \emph {et~al.}(2016)\citenamefont
  {Bzdusek}, \citenamefont {Wu}, \citenamefont {Ruegg}, \citenamefont
  {Sigrist}, \citenamefont {Soluyanov} \emph {et~al.}}]{21bzduvsek2016nodal}%
  \BibitemOpen
  \bibfield  {author} {\bibinfo {author} {\bibfnamefont {T.}~\bibnamefont
  {Bzdusek}}, \bibinfo {author} {\bibfnamefont {Q.}~\bibnamefont {Wu}},
  \bibinfo {author} {\bibfnamefont {A.}~\bibnamefont {Ruegg}}, \bibinfo
  {author} {\bibfnamefont {M.}~\bibnamefont {Sigrist}}, \bibinfo {author}
  {\bibfnamefont {A.}~\bibnamefont {Soluyanov}},  \emph {et~al.},\ }\href@noop
  {} {\bibfield  {journal} {\bibinfo  {journal} {Nature}\ }\textbf {\bibinfo
  {volume} {538}},\ \bibinfo {pages} {75} (\bibinfo {year} {2016})}\BibitemShut
  {NoStop}%
\bibitem [{\citenamefont {Wang}\ \emph
  {et~al.}(2017{\natexlab{a}})\citenamefont {Wang}, \citenamefont {Liu},
  \citenamefont {Yu}, \citenamefont {Sheng},\ and\ \citenamefont
  {Yang}}]{22wang2017hourglass}%
  \BibitemOpen
  \bibfield  {author} {\bibinfo {author} {\bibfnamefont {S.-S.}\ \bibnamefont
  {Wang}}, \bibinfo {author} {\bibfnamefont {Y.}~\bibnamefont {Liu}}, \bibinfo
  {author} {\bibfnamefont {Z.-M.}\ \bibnamefont {Yu}}, \bibinfo {author}
  {\bibfnamefont {X.-L.}\ \bibnamefont {Sheng}}, \ and\ \bibinfo {author}
  {\bibfnamefont {S.~A.}\ \bibnamefont {Yang}},\ }\href@noop {} {\bibfield
  {journal} {\bibinfo  {journal} {Nat. Commun.}\ }\textbf {\bibinfo {volume}
  {8}},\ \bibinfo {pages} {1844} (\bibinfo {year}
  {2017}{\natexlab{a}})}\BibitemShut {NoStop}%
\bibitem [{\citenamefont {Yan}\ \emph {et~al.}(2017)\citenamefont {Yan},
  \citenamefont {Bi}, \citenamefont {Shen}, \citenamefont {Lu}, \citenamefont
  {Zhang},\ and\ \citenamefont {Wang}}]{23yan2017nodal}%
  \BibitemOpen
  \bibfield  {author} {\bibinfo {author} {\bibfnamefont {Z.}~\bibnamefont
  {Yan}}, \bibinfo {author} {\bibfnamefont {R.}~\bibnamefont {Bi}}, \bibinfo
  {author} {\bibfnamefont {H.}~\bibnamefont {Shen}}, \bibinfo {author}
  {\bibfnamefont {L.}~\bibnamefont {Lu}}, \bibinfo {author} {\bibfnamefont
  {S.-C.}\ \bibnamefont {Zhang}}, \ and\ \bibinfo {author} {\bibfnamefont
  {Z.}~\bibnamefont {Wang}},\ }\href@noop {} {\bibfield  {journal} {\bibinfo
  {journal} {Phys. Rev. B}\ }\textbf {\bibinfo {volume} {96}},\ \bibinfo
  {pages} {041103} (\bibinfo {year} {2017})}\BibitemShut {NoStop}%
\bibitem [{\citenamefont {Chen}\ \emph {et~al.}(2017)\citenamefont {Chen},
  \citenamefont {Lu},\ and\ \citenamefont {Hou}}]{24chen2017topological}%
  \BibitemOpen
  \bibfield  {author} {\bibinfo {author} {\bibfnamefont {W.}~\bibnamefont
  {Chen}}, \bibinfo {author} {\bibfnamefont {H.-Z.}\ \bibnamefont {Lu}}, \ and\
  \bibinfo {author} {\bibfnamefont {J.-M.}\ \bibnamefont {Hou}},\ }\href@noop
  {} {\bibfield  {journal} {\bibinfo  {journal} {Phys. Rev. B}\ }\textbf
  {\bibinfo {volume} {96}},\ \bibinfo {pages} {041102} (\bibinfo {year}
  {2017})}\BibitemShut {NoStop}%
\bibitem [{\citenamefont {Fu}\ \emph {et~al.}(2018)\citenamefont {Fu},
  \citenamefont {Fan}, \citenamefont {Ma}, \citenamefont {Liu},\ and\
  \citenamefont {Yao}}]{25fu2018hourglasslike}%
  \BibitemOpen
  \bibfield  {author} {\bibinfo {author} {\bibfnamefont {B.}~\bibnamefont
  {Fu}}, \bibinfo {author} {\bibfnamefont {X.}~\bibnamefont {Fan}}, \bibinfo
  {author} {\bibfnamefont {D.}~\bibnamefont {Ma}}, \bibinfo {author}
  {\bibfnamefont {C.-C.}\ \bibnamefont {Liu}}, \ and\ \bibinfo {author}
  {\bibfnamefont {Y.}~\bibnamefont {Yao}},\ }\href@noop {} {\bibfield
  {journal} {\bibinfo  {journal} {Phys. Rev. B}\ }\textbf {\bibinfo {volume}
  {98}},\ \bibinfo {pages} {075146} (\bibinfo {year} {2018})}\BibitemShut
  {NoStop}%
\bibitem [{\citenamefont {Singh}\ \emph {et~al.}(2018)\citenamefont {Singh},
  \citenamefont {Ghosh}, \citenamefont {Su}, \citenamefont {Lin}, \citenamefont
  {Agarwal},\ and\ \citenamefont {Bansil}}]{26singh2018topological}%
  \BibitemOpen
  \bibfield  {author} {\bibinfo {author} {\bibfnamefont {B.}~\bibnamefont
  {Singh}}, \bibinfo {author} {\bibfnamefont {B.}~\bibnamefont {Ghosh}},
  \bibinfo {author} {\bibfnamefont {C.}~\bibnamefont {Su}}, \bibinfo {author}
  {\bibfnamefont {H.}~\bibnamefont {Lin}}, \bibinfo {author} {\bibfnamefont
  {A.}~\bibnamefont {Agarwal}}, \ and\ \bibinfo {author} {\bibfnamefont
  {A.}~\bibnamefont {Bansil}},\ }\href@noop {} {\bibfield  {journal} {\bibinfo
  {journal} {Phys. Rev. Lett.}\ }\textbf {\bibinfo {volume} {121}},\ \bibinfo
  {pages} {226401} (\bibinfo {year} {2018})}\BibitemShut {NoStop}%
\bibitem [{\citenamefont {Feng}\ \emph {et~al.}(2018)\citenamefont {Feng},
  \citenamefont {Yue}, \citenamefont {Song}, \citenamefont {Wu},\ and\
  \citenamefont {Wen}}]{27feng2018topological}%
  \BibitemOpen
  \bibfield  {author} {\bibinfo {author} {\bibfnamefont {X.}~\bibnamefont
  {Feng}}, \bibinfo {author} {\bibfnamefont {C.}~\bibnamefont {Yue}}, \bibinfo
  {author} {\bibfnamefont {Z.}~\bibnamefont {Song}}, \bibinfo {author}
  {\bibfnamefont {Q.}~\bibnamefont {Wu}}, \ and\ \bibinfo {author}
  {\bibfnamefont {B.}~\bibnamefont {Wen}},\ }\href@noop {} {\bibfield
  {journal} {\bibinfo  {journal} {Phys. Rev. Materials}\ }\textbf {\bibinfo
  {volume} {2}},\ \bibinfo {pages} {014202} (\bibinfo {year}
  {2018})}\BibitemShut {NoStop}%
\bibitem [{\citenamefont {Wang}\ \emph {et~al.}(2018)\citenamefont {Wang},
  \citenamefont {Nie}, \citenamefont {Weng}, \citenamefont {Kawazoe},\ and\
  \citenamefont {Chen}}]{28wang2018topological}%
  \BibitemOpen
  \bibfield  {author} {\bibinfo {author} {\bibfnamefont {J.-T.}\ \bibnamefont
  {Wang}}, \bibinfo {author} {\bibfnamefont {S.}~\bibnamefont {Nie}}, \bibinfo
  {author} {\bibfnamefont {H.}~\bibnamefont {Weng}}, \bibinfo {author}
  {\bibfnamefont {Y.}~\bibnamefont {Kawazoe}}, \ and\ \bibinfo {author}
  {\bibfnamefont {C.}~\bibnamefont {Chen}},\ }\href@noop {} {\bibfield
  {journal} {\bibinfo  {journal} {Phys. Rev. Lett.}\ }\textbf {\bibinfo
  {volume} {120}},\ \bibinfo {pages} {026402} (\bibinfo {year}
  {2018})}\BibitemShut {NoStop}%
\bibitem [{\citenamefont {Ezawa}(2017)}]{29ezawa2017topological}%
  \BibitemOpen
  \bibfield  {author} {\bibinfo {author} {\bibfnamefont {M.}~\bibnamefont
  {Ezawa}},\ }\href@noop {} {\bibfield  {journal} {\bibinfo  {journal} {Phys.
  Rev. B}\ }\textbf {\bibinfo {volume} {96}},\ \bibinfo {pages} {041202}
  (\bibinfo {year} {2017})}\BibitemShut {NoStop}%
\bibitem [{\citenamefont {Bi}\ \emph {et~al.}(2017)\citenamefont {Bi},
  \citenamefont {Yan}, \citenamefont {Lu},\ and\ \citenamefont
  {Wang}}]{30bi2017nodal}%
  \BibitemOpen
  \bibfield  {author} {\bibinfo {author} {\bibfnamefont {R.}~\bibnamefont
  {Bi}}, \bibinfo {author} {\bibfnamefont {Z.}~\bibnamefont {Yan}}, \bibinfo
  {author} {\bibfnamefont {L.}~\bibnamefont {Lu}}, \ and\ \bibinfo {author}
  {\bibfnamefont {Z.}~\bibnamefont {Wang}},\ }\href@noop {} {\bibfield
  {journal} {\bibinfo  {journal} {Phys. Rev. B}\ }\textbf {\bibinfo {volume}
  {96}},\ \bibinfo {pages} {201305} (\bibinfo {year} {2017})}\BibitemShut
  {NoStop}%
\bibitem [{\citenamefont {Kane}\ and\ \citenamefont
  {Mele}(2005)}]{31kane2005quantum}%
  \BibitemOpen
  \bibfield  {author} {\bibinfo {author} {\bibfnamefont {C.~L.}\ \bibnamefont
  {Kane}}\ and\ \bibinfo {author} {\bibfnamefont {E.~J.}\ \bibnamefont
  {Mele}},\ }\href@noop {} {\bibfield  {journal} {\bibinfo  {journal} {Phys.
  Rev. Lett.}\ }\textbf {\bibinfo {volume} {95}},\ \bibinfo {pages} {226801}
  (\bibinfo {year} {2005})}\BibitemShut {NoStop}%
\bibitem [{\citenamefont {Bernevig}\ \emph {et~al.}(2006)\citenamefont
  {Bernevig}, \citenamefont {Hughes},\ and\ \citenamefont
  {Zhang}}]{32bernevig2006quantum}%
  \BibitemOpen
  \bibfield  {author} {\bibinfo {author} {\bibfnamefont {B.~A.}\ \bibnamefont
  {Bernevig}}, \bibinfo {author} {\bibfnamefont {T.~L.}\ \bibnamefont
  {Hughes}}, \ and\ \bibinfo {author} {\bibfnamefont {S.-C.}\ \bibnamefont
  {Zhang}},\ }\href@noop {} {\bibfield  {journal} {\bibinfo  {journal}
  {Science}\ }\textbf {\bibinfo {volume} {314}},\ \bibinfo {pages} {1757}
  (\bibinfo {year} {2006})}\BibitemShut {NoStop}%
\bibitem [{\citenamefont {Wan}\ \emph {et~al.}(2011)\citenamefont {Wan},
  \citenamefont {Turner}, \citenamefont {Vishwanath},\ and\ \citenamefont
  {Savrasov}}]{33wan2011topological}%
  \BibitemOpen
  \bibfield  {author} {\bibinfo {author} {\bibfnamefont {X.}~\bibnamefont
  {Wan}}, \bibinfo {author} {\bibfnamefont {A.~M.}\ \bibnamefont {Turner}},
  \bibinfo {author} {\bibfnamefont {A.}~\bibnamefont {Vishwanath}}, \ and\
  \bibinfo {author} {\bibfnamefont {S.~Y.}\ \bibnamefont {Savrasov}},\
  }\href@noop {} {\bibfield  {journal} {\bibinfo  {journal} {Phys. Rev. B}\
  }\textbf {\bibinfo {volume} {83}},\ \bibinfo {pages} {205101} (\bibinfo
  {year} {2011})}\BibitemShut {NoStop}%
\bibitem [{\citenamefont {Young}\ \emph {et~al.}(2012)\citenamefont {Young},
  \citenamefont {Zaheer}, \citenamefont {Teo}, \citenamefont {Kane},
  \citenamefont {Mele},\ and\ \citenamefont {Rappe}}]{34young2012dirac}%
  \BibitemOpen
  \bibfield  {author} {\bibinfo {author} {\bibfnamefont {S.~M.}\ \bibnamefont
  {Young}}, \bibinfo {author} {\bibfnamefont {S.}~\bibnamefont {Zaheer}},
  \bibinfo {author} {\bibfnamefont {J.~C.~Y.}\ \bibnamefont {Teo}}, \bibinfo
  {author} {\bibfnamefont {C.~L.}\ \bibnamefont {Kane}}, \bibinfo {author}
  {\bibfnamefont {E.~J.}\ \bibnamefont {Mele}}, \ and\ \bibinfo {author}
  {\bibfnamefont {A.~M.}\ \bibnamefont {Rappe}},\ }\href@noop {} {\bibfield
  {journal} {\bibinfo  {journal} {Phys. Rev. Lett.}\ }\textbf {\bibinfo
  {volume} {108}},\ \bibinfo {pages} {140405} (\bibinfo {year}
  {2012})}\BibitemShut {NoStop}%
\bibitem [{\citenamefont {Yang}\ and\ \citenamefont
  {Nagaosa}(2014)}]{35yang2014classification}%
  \BibitemOpen
  \bibfield  {author} {\bibinfo {author} {\bibfnamefont {B.-J.}\ \bibnamefont
  {Yang}}\ and\ \bibinfo {author} {\bibfnamefont {N.}~\bibnamefont {Nagaosa}},\
  }\href@noop {} {\bibfield  {journal} {\bibinfo  {journal} {Nat. Commun.}\
  }\textbf {\bibinfo {volume} {5}} (\bibinfo {year} {2014})}\BibitemShut
  {NoStop}%
\bibitem [{\citenamefont {Wang}\ \emph {et~al.}(2012)\citenamefont {Wang},
  \citenamefont {Sun}, \citenamefont {Chen}, \citenamefont {Franchini},
  \citenamefont {Xu}, \citenamefont {Weng}, \citenamefont {Dai},\ and\
  \citenamefont {Fang}}]{36wang2012dirac}%
  \BibitemOpen
  \bibfield  {author} {\bibinfo {author} {\bibfnamefont {Z.}~\bibnamefont
  {Wang}}, \bibinfo {author} {\bibfnamefont {Y.}~\bibnamefont {Sun}}, \bibinfo
  {author} {\bibfnamefont {X.-Q.}\ \bibnamefont {Chen}}, \bibinfo {author}
  {\bibfnamefont {C.}~\bibnamefont {Franchini}}, \bibinfo {author}
  {\bibfnamefont {G.}~\bibnamefont {Xu}}, \bibinfo {author} {\bibfnamefont
  {H.}~\bibnamefont {Weng}}, \bibinfo {author} {\bibfnamefont {X.}~\bibnamefont
  {Dai}}, \ and\ \bibinfo {author} {\bibfnamefont {Z.}~\bibnamefont {Fang}},\
  }\href@noop {} {\bibfield  {journal} {\bibinfo  {journal} {Phys. Rev. B}\
  }\textbf {\bibinfo {volume} {85}},\ \bibinfo {pages} {195320} (\bibinfo
  {year} {2012})}\BibitemShut {NoStop}%
\bibitem [{\citenamefont {Wang}\ \emph {et~al.}(2013)\citenamefont {Wang},
  \citenamefont {Weng}, \citenamefont {Wu}, \citenamefont {Dai},\ and\
  \citenamefont {Fang}}]{37wang2013three}%
  \BibitemOpen
  \bibfield  {author} {\bibinfo {author} {\bibfnamefont {Z.}~\bibnamefont
  {Wang}}, \bibinfo {author} {\bibfnamefont {H.}~\bibnamefont {Weng}}, \bibinfo
  {author} {\bibfnamefont {Q.}~\bibnamefont {Wu}}, \bibinfo {author}
  {\bibfnamefont {X.}~\bibnamefont {Dai}}, \ and\ \bibinfo {author}
  {\bibfnamefont {Z.}~\bibnamefont {Fang}},\ }\href@noop {} {\bibfield
  {journal} {\bibinfo  {journal} {Phys. Rev. B}\ }\textbf {\bibinfo {volume}
  {88}},\ \bibinfo {pages} {125427} (\bibinfo {year} {2013})}\BibitemShut
  {NoStop}%
\bibitem [{\citenamefont {Chang}\ \emph
  {et~al.}(2017{\natexlab{a}})\citenamefont {Chang}, \citenamefont {Xu},
  \citenamefont {Sanchez}, \citenamefont {Tsai}, \citenamefont {Huang},
  \citenamefont {Chang}, \citenamefont {Hsu}, \citenamefont {Bian},
  \citenamefont {Belopolski}, \citenamefont {Yu}, \citenamefont {Yang},
  \citenamefont {Neupert}, \citenamefont {Jeng}, \citenamefont {Lin},\ and\
  \citenamefont {Hasan}}]{38chang2017type}%
  \BibitemOpen
  \bibfield  {author} {\bibinfo {author} {\bibfnamefont {T.-R.}\ \bibnamefont
  {Chang}}, \bibinfo {author} {\bibfnamefont {S.-Y.}\ \bibnamefont {Xu}},
  \bibinfo {author} {\bibfnamefont {D.~S.}\ \bibnamefont {Sanchez}}, \bibinfo
  {author} {\bibfnamefont {W.-F.}\ \bibnamefont {Tsai}}, \bibinfo {author}
  {\bibfnamefont {S.-M.}\ \bibnamefont {Huang}}, \bibinfo {author}
  {\bibfnamefont {G.}~\bibnamefont {Chang}}, \bibinfo {author} {\bibfnamefont
  {C.-H.}\ \bibnamefont {Hsu}}, \bibinfo {author} {\bibfnamefont
  {G.}~\bibnamefont {Bian}}, \bibinfo {author} {\bibfnamefont {I.}~\bibnamefont
  {Belopolski}}, \bibinfo {author} {\bibfnamefont {Z.-M.}\ \bibnamefont {Yu}},
  \bibinfo {author} {\bibfnamefont {S.~A.}\ \bibnamefont {Yang}}, \bibinfo
  {author} {\bibfnamefont {T.}~\bibnamefont {Neupert}}, \bibinfo {author}
  {\bibfnamefont {H.-T.}\ \bibnamefont {Jeng}}, \bibinfo {author}
  {\bibfnamefont {H.}~\bibnamefont {Lin}}, \ and\ \bibinfo {author}
  {\bibfnamefont {M.~Z.}\ \bibnamefont {Hasan}},\ }\href@noop {} {\bibfield
  {journal} {\bibinfo  {journal} {Phys. Rev. Lett.}\ }\textbf {\bibinfo
  {volume} {119}},\ \bibinfo {pages} {026404} (\bibinfo {year}
  {2017}{\natexlab{a}})}\BibitemShut {NoStop}%
\bibitem [{\citenamefont {Xu}\ \emph {et~al.}(2011)\citenamefont {Xu},
  \citenamefont {Weng}, \citenamefont {Wang}, \citenamefont {Dai},\ and\
  \citenamefont {Fang}}]{39xu2011chern}%
  \BibitemOpen
  \bibfield  {author} {\bibinfo {author} {\bibfnamefont {G.}~\bibnamefont
  {Xu}}, \bibinfo {author} {\bibfnamefont {H.}~\bibnamefont {Weng}}, \bibinfo
  {author} {\bibfnamefont {Z.}~\bibnamefont {Wang}}, \bibinfo {author}
  {\bibfnamefont {X.}~\bibnamefont {Dai}}, \ and\ \bibinfo {author}
  {\bibfnamefont {Z.}~\bibnamefont {Fang}},\ }\href@noop {} {\bibfield
  {journal} {\bibinfo  {journal} {Phys. Rev. Lett.}\ }\textbf {\bibinfo
  {volume} {107}},\ \bibinfo {pages} {186806} (\bibinfo {year}
  {2011})}\BibitemShut {NoStop}%
\bibitem [{\citenamefont {Weng}\ \emph
  {et~al.}(2015{\natexlab{b}})\citenamefont {Weng}, \citenamefont {Fang},
  \citenamefont {Fang}, \citenamefont {Bernevig},\ and\ \citenamefont
  {Dai}}]{40weng2015weyl}%
  \BibitemOpen
  \bibfield  {author} {\bibinfo {author} {\bibfnamefont {H.}~\bibnamefont
  {Weng}}, \bibinfo {author} {\bibfnamefont {C.}~\bibnamefont {Fang}}, \bibinfo
  {author} {\bibfnamefont {Z.}~\bibnamefont {Fang}}, \bibinfo {author}
  {\bibfnamefont {B.~A.}\ \bibnamefont {Bernevig}}, \ and\ \bibinfo {author}
  {\bibfnamefont {X.}~\bibnamefont {Dai}},\ }\href@noop {} {\bibfield
  {journal} {\bibinfo  {journal} {Phys. Rev. X}\ }\textbf {\bibinfo {volume}
  {5}},\ \bibinfo {pages} {011029} (\bibinfo {year}
  {2015}{\natexlab{b}})}\BibitemShut {NoStop}%
\bibitem [{\citenamefont {Soluyanov}\ \emph {et~al.}(2015)\citenamefont
  {Soluyanov}, \citenamefont {Gresch}, \citenamefont {Wang}, \citenamefont
  {Wu}, \citenamefont {Troyer}, \citenamefont {Dai},\ and\ \citenamefont
  {Bernevig}}]{41soluyanov2015type}%
  \BibitemOpen
  \bibfield  {author} {\bibinfo {author} {\bibfnamefont {A.~A.}\ \bibnamefont
  {Soluyanov}}, \bibinfo {author} {\bibfnamefont {D.}~\bibnamefont {Gresch}},
  \bibinfo {author} {\bibfnamefont {Z.}~\bibnamefont {Wang}}, \bibinfo {author}
  {\bibfnamefont {Q.}~\bibnamefont {Wu}}, \bibinfo {author} {\bibfnamefont
  {M.}~\bibnamefont {Troyer}}, \bibinfo {author} {\bibfnamefont
  {X.}~\bibnamefont {Dai}}, \ and\ \bibinfo {author} {\bibfnamefont {B.~A.}\
  \bibnamefont {Bernevig}},\ }\href@noop {} {\bibfield  {journal} {\bibinfo
  {journal} {Nature}\ }\textbf {\bibinfo {volume} {527}},\ \bibinfo {pages}
  {495} (\bibinfo {year} {2015})}\BibitemShut {NoStop}%
\bibitem [{\citenamefont {Ruan}\ \emph {et~al.}(2016)\citenamefont {Ruan},
  \citenamefont {Jian}, \citenamefont {Zhang}, \citenamefont {Yao},
  \citenamefont {Zhang}, \citenamefont {Zhang},\ and\ \citenamefont
  {Xing}}]{42ruan2016ideal}%
  \BibitemOpen
  \bibfield  {author} {\bibinfo {author} {\bibfnamefont {J.}~\bibnamefont
  {Ruan}}, \bibinfo {author} {\bibfnamefont {S.-K.}\ \bibnamefont {Jian}},
  \bibinfo {author} {\bibfnamefont {D.}~\bibnamefont {Zhang}}, \bibinfo
  {author} {\bibfnamefont {H.}~\bibnamefont {Yao}}, \bibinfo {author}
  {\bibfnamefont {H.}~\bibnamefont {Zhang}}, \bibinfo {author} {\bibfnamefont
  {S.-C.}\ \bibnamefont {Zhang}}, \ and\ \bibinfo {author} {\bibfnamefont
  {D.}~\bibnamefont {Xing}},\ }\href@noop {} {\bibfield  {journal} {\bibinfo
  {journal} {Phys. Rev. Lett.}\ }\textbf {\bibinfo {volume} {116}},\ \bibinfo
  {pages} {226801} (\bibinfo {year} {2016})}\BibitemShut {NoStop}%
\bibitem [{\citenamefont {Aut\`es}\ \emph {et~al.}(2016)\citenamefont
  {Aut\`es}, \citenamefont {Gresch}, \citenamefont {Troyer}, \citenamefont
  {Soluyanov},\ and\ \citenamefont {Yazyev}}]{43autes2016robust}%
  \BibitemOpen
  \bibfield  {author} {\bibinfo {author} {\bibfnamefont {G.}~\bibnamefont
  {Aut\`es}}, \bibinfo {author} {\bibfnamefont {D.}~\bibnamefont {Gresch}},
  \bibinfo {author} {\bibfnamefont {M.}~\bibnamefont {Troyer}}, \bibinfo
  {author} {\bibfnamefont {A.~A.}\ \bibnamefont {Soluyanov}}, \ and\ \bibinfo
  {author} {\bibfnamefont {O.~V.}\ \bibnamefont {Yazyev}},\ }\href@noop {}
  {\bibfield  {journal} {\bibinfo  {journal} {Phys. Rev. Lett.}\ }\textbf
  {\bibinfo {volume} {117}},\ \bibinfo {pages} {066402} (\bibinfo {year}
  {2016})}\BibitemShut {NoStop}%
\bibitem [{\citenamefont {Zhu}\ \emph {et~al.}(2016)\citenamefont {Zhu},
  \citenamefont {Winkler}, \citenamefont {Wu}, \citenamefont {Li},\ and\
  \citenamefont {Soluyanov}}]{44zhu2016triple}%
  \BibitemOpen
  \bibfield  {author} {\bibinfo {author} {\bibfnamefont {Z.}~\bibnamefont
  {Zhu}}, \bibinfo {author} {\bibfnamefont {G.~W.}\ \bibnamefont {Winkler}},
  \bibinfo {author} {\bibfnamefont {Q.}~\bibnamefont {Wu}}, \bibinfo {author}
  {\bibfnamefont {J.}~\bibnamefont {Li}}, \ and\ \bibinfo {author}
  {\bibfnamefont {A.~A.}\ \bibnamefont {Soluyanov}},\ }\href@noop {} {\bibfield
   {journal} {\bibinfo  {journal} {Phys. Rev. X}\ }\textbf {\bibinfo {volume}
  {6}},\ \bibinfo {pages} {031003} (\bibinfo {year} {2016})}\BibitemShut
  {NoStop}%
\bibitem [{\citenamefont {Wang}\ \emph
  {et~al.}(2017{\natexlab{b}})\citenamefont {Wang}, \citenamefont {Sui},
  \citenamefont {Shi}, \citenamefont {Pan}, \citenamefont {Zhang},
  \citenamefont {Liu}, \citenamefont {Wei}, \citenamefont {Yan},\ and\
  \citenamefont {Huang}}]{45wang2017prediction}%
  \BibitemOpen
  \bibfield  {author} {\bibinfo {author} {\bibfnamefont {J.}~\bibnamefont
  {Wang}}, \bibinfo {author} {\bibfnamefont {X.}~\bibnamefont {Sui}}, \bibinfo
  {author} {\bibfnamefont {W.}~\bibnamefont {Shi}}, \bibinfo {author}
  {\bibfnamefont {J.}~\bibnamefont {Pan}}, \bibinfo {author} {\bibfnamefont
  {S.}~\bibnamefont {Zhang}}, \bibinfo {author} {\bibfnamefont
  {F.}~\bibnamefont {Liu}}, \bibinfo {author} {\bibfnamefont {S.-H.}\
  \bibnamefont {Wei}}, \bibinfo {author} {\bibfnamefont {Q.}~\bibnamefont
  {Yan}}, \ and\ \bibinfo {author} {\bibfnamefont {B.}~\bibnamefont {Huang}},\
  }\href@noop {} {\bibfield  {journal} {\bibinfo  {journal} {Phys. Rev. Lett.}\
  }\textbf {\bibinfo {volume} {119}},\ \bibinfo {pages} {256402} (\bibinfo
  {year} {2017}{\natexlab{b}})}\BibitemShut {NoStop}%
\bibitem [{\citenamefont {Rhim}\ and\ \citenamefont
  {Kim}(2015)}]{46rhim2015landau}%
  \BibitemOpen
  \bibfield  {author} {\bibinfo {author} {\bibfnamefont {J.-W.}\ \bibnamefont
  {Rhim}}\ and\ \bibinfo {author} {\bibfnamefont {Y.~B.}\ \bibnamefont {Kim}},\
  }\href@noop {} {\bibfield  {journal} {\bibinfo  {journal} {Phys. Rev. B}\
  }\textbf {\bibinfo {volume} {92}},\ \bibinfo {pages} {045126} (\bibinfo
  {year} {2015})}\BibitemShut {NoStop}%
\bibitem [{\citenamefont {Huh}\ \emph {et~al.}(2016)\citenamefont {Huh},
  \citenamefont {Moon},\ and\ \citenamefont {Kim}}]{47huh2016long}%
  \BibitemOpen
  \bibfield  {author} {\bibinfo {author} {\bibfnamefont {Y.}~\bibnamefont
  {Huh}}, \bibinfo {author} {\bibfnamefont {E.-G.}\ \bibnamefont {Moon}}, \
  and\ \bibinfo {author} {\bibfnamefont {Y.~B.}\ \bibnamefont {Kim}},\
  }\href@noop {} {\bibfield  {journal} {\bibinfo  {journal} {Phys. Rev. B}\
  }\textbf {\bibinfo {volume} {93}},\ \bibinfo {pages} {035138} (\bibinfo
  {year} {2016})}\BibitemShut {NoStop}%
\bibitem [{\citenamefont {Hepworth}\ \emph {et~al.}(1957)\citenamefont
  {Hepworth}, \citenamefont {Jack}, \citenamefont {Peacock},\ and\
  \citenamefont {Westland}}]{48hepworth1957crystalPdF3}%
  \BibitemOpen
  \bibfield  {author} {\bibinfo {author} {\bibfnamefont {M.}~\bibnamefont
  {Hepworth}}, \bibinfo {author} {\bibfnamefont {K.}~\bibnamefont {Jack}},
  \bibinfo {author} {\bibfnamefont {R.}~\bibnamefont {Peacock}}, \ and\
  \bibinfo {author} {\bibfnamefont {G.}~\bibnamefont {Westland}},\ }\href@noop
  {} {\bibfield  {journal} {\bibinfo  {journal} {Acta Cryst.}\ }\textbf
  {\bibinfo {volume} {10}},\ \bibinfo {pages} {63} (\bibinfo {year}
  {1957})}\BibitemShut {NoStop}%
\bibitem [{\citenamefont {M{\"u}ller}\ and\ \citenamefont
  {Serafin}(1987)}]{49muller1987kristallstruktur}%
  \BibitemOpen
  \bibfield  {author} {\bibinfo {author} {\bibfnamefont {B.~G.}\ \bibnamefont
  {M{\"u}ller}}\ and\ \bibinfo {author} {\bibfnamefont {M.}~\bibnamefont
  {Serafin}},\ }\href@noop {} {\bibfield  {journal} {\bibinfo  {journal} {Z.
  Naturforsch. B}\ }\textbf {\bibinfo {volume} {42}},\ \bibinfo {pages} {1102}
  (\bibinfo {year} {1987})}\BibitemShut {NoStop}%
\bibitem [{\citenamefont {Effenberger}\ \emph {et~al.}(1981)\citenamefont
  {Effenberger}, \citenamefont {Mereiter},\ and\ \citenamefont
  {Zemann}}]{50effenberger1981crystal}%
  \BibitemOpen
  \bibfield  {author} {\bibinfo {author} {\bibfnamefont {H.}~\bibnamefont
  {Effenberger}}, \bibinfo {author} {\bibfnamefont {K.}~\bibnamefont
  {Mereiter}}, \ and\ \bibinfo {author} {\bibfnamefont {J.}~\bibnamefont
  {Zemann}},\ }\href@noop {} {\bibfield  {journal} {\bibinfo  {journal} {Z.
  Krist.-Cryst. Mater.}\ }\textbf {\bibinfo {volume} {156}},\ \bibinfo {pages}
  {233} (\bibinfo {year} {1981})}\BibitemShut {NoStop}%
\bibitem [{\citenamefont {Huber}\ and\ \citenamefont
  {Deiseroth}(1995)}]{51huber1995crystal}%
  \BibitemOpen
  \bibfield  {author} {\bibinfo {author} {\bibfnamefont {M.}~\bibnamefont
  {Huber}}\ and\ \bibinfo {author} {\bibfnamefont {H.}~\bibnamefont
  {Deiseroth}},\ }\href@noop {} {\bibfield  {journal} {\bibinfo  {journal} {Z.
  Krist.}\ }\textbf {\bibinfo {volume} {210}},\ \bibinfo {pages} {685}
  (\bibinfo {year} {1995})}\BibitemShut {NoStop}%
\bibitem [{\citenamefont {Moreno}\ \emph {et~al.}(2008)\citenamefont {Moreno},
  \citenamefont {Valencia}, \citenamefont {T{\'e}llez}, \citenamefont
  {Mart{\'\i}nez}, \citenamefont {Roa-Rojas}, \citenamefont {Fajardo} \emph
  {et~al.}}]{52moreno2008preparation}%
  \BibitemOpen
  \bibfield  {author} {\bibinfo {author} {\bibfnamefont {L.}~\bibnamefont
  {Moreno}}, \bibinfo {author} {\bibfnamefont {J.}~\bibnamefont {Valencia}},
  \bibinfo {author} {\bibfnamefont {D.~L.}\ \bibnamefont {T{\'e}llez}},
  \bibinfo {author} {\bibfnamefont {M.}~\bibnamefont {Mart{\'\i}nez}}, \bibinfo
  {author} {\bibfnamefont {J.}~\bibnamefont {Roa-Rojas}}, \bibinfo {author}
  {\bibfnamefont {F.}~\bibnamefont {Fajardo}},  \emph {et~al.},\ }\href@noop {}
  {\bibfield  {journal} {\bibinfo  {journal} {J Magn. Magn. Mater.}\ }\textbf
  {\bibinfo {volume} {320}},\ \bibinfo {pages} {e19} (\bibinfo {year}
  {2008})}\BibitemShut {NoStop}%
\bibitem [{\citenamefont {Garcia-Munoz}\ \emph {et~al.}(1992)\citenamefont
  {Garcia-Munoz}, \citenamefont {Rodriguez-Carvajal}, \citenamefont {Lacorre},\
  and\ \citenamefont {Torrance}}]{53garcia1992neutron}%
  \BibitemOpen
  \bibfield  {author} {\bibinfo {author} {\bibfnamefont {J.}~\bibnamefont
  {Garcia-Munoz}}, \bibinfo {author} {\bibfnamefont {J.}~\bibnamefont
  {Rodriguez-Carvajal}}, \bibinfo {author} {\bibfnamefont {P.}~\bibnamefont
  {Lacorre}}, \ and\ \bibinfo {author} {\bibfnamefont {J.}~\bibnamefont
  {Torrance}},\ }\href@noop {} {\bibfield  {journal} {\bibinfo  {journal}
  {Phys. Rev. B}\ }\textbf {\bibinfo {volume} {46}},\ \bibinfo {pages} {4414}
  (\bibinfo {year} {1992})}\BibitemShut {NoStop}%
\bibitem [{54()}]{54}%
  \BibitemOpen
  \href@noop {} {\bibinfo  {journal} {See Supplemental Material for the
  computational methods, the symmetry operators analysis, the tight-binding
  model, and the supplementary figures of candidates}\ }\BibitemShut {NoStop}%
\bibitem [{\citenamefont {Jain}\ \emph {et~al.}(2013)\citenamefont {Jain},
  \citenamefont {Ong}, \citenamefont {Hautier}, \citenamefont {Chen},
  \citenamefont {Richards}, \citenamefont {Dacek}, \citenamefont {Cholia},
  \citenamefont {Gunter}, \citenamefont {Skinner}, \citenamefont {Ceder} \emph
  {et~al.}}]{55jain2013commentary}%
  \BibitemOpen
\bibfield  {journal} {  }\bibfield  {author} {\bibinfo {author} {\bibfnamefont
  {A.}~\bibnamefont {Jain}}, \bibinfo {author} {\bibfnamefont {S.~P.}\
  \bibnamefont {Ong}}, \bibinfo {author} {\bibfnamefont {G.}~\bibnamefont
  {Hautier}}, \bibinfo {author} {\bibfnamefont {W.}~\bibnamefont {Chen}},
  \bibinfo {author} {\bibfnamefont {W.~D.}\ \bibnamefont {Richards}}, \bibinfo
  {author} {\bibfnamefont {S.}~\bibnamefont {Dacek}}, \bibinfo {author}
  {\bibfnamefont {S.}~\bibnamefont {Cholia}}, \bibinfo {author} {\bibfnamefont
  {D.}~\bibnamefont {Gunter}}, \bibinfo {author} {\bibfnamefont
  {D.}~\bibnamefont {Skinner}}, \bibinfo {author} {\bibfnamefont
  {G.}~\bibnamefont {Ceder}},  \emph {et~al.},\ }\href@noop {} {\bibfield
  {journal} {\bibinfo  {journal} {Apl Mater.}\ }\textbf {\bibinfo {volume}
  {1}},\ \bibinfo {pages} {011002} (\bibinfo {year} {2013})}\BibitemShut
  {NoStop}%
\bibitem [{\citenamefont {Fu}(2011)}]{56fu2011topological}%
  \BibitemOpen
  \bibfield  {author} {\bibinfo {author} {\bibfnamefont {L.}~\bibnamefont
  {Fu}},\ }\href@noop {} {\bibfield  {journal} {\bibinfo  {journal} {Phys. Rev.
  Lett.}\ }\textbf {\bibinfo {volume} {106}},\ \bibinfo {pages} {106802}
  (\bibinfo {year} {2011})}\BibitemShut {NoStop}%
\bibitem [{\citenamefont {Young}\ and\ \citenamefont
  {Kane}(2015)}]{57young2015dirac}%
  \BibitemOpen
  \bibfield  {author} {\bibinfo {author} {\bibfnamefont {S.~M.}\ \bibnamefont
  {Young}}\ and\ \bibinfo {author} {\bibfnamefont {C.~L.}\ \bibnamefont
  {Kane}},\ }\href@noop {} {\bibfield  {journal} {\bibinfo  {journal} {Phys.
  Rev. Lett.}\ }\textbf {\bibinfo {volume} {115}},\ \bibinfo {pages} {126803}
  (\bibinfo {year} {2015})}\BibitemShut {NoStop}%
\bibitem [{\citenamefont {Wang}\ \emph {et~al.}(2016)\citenamefont {Wang},
  \citenamefont {Vergniory}, \citenamefont {Kushwaha}, \citenamefont
  {Hirschberger}, \citenamefont {Chulkov}, \citenamefont {Ernst}, \citenamefont
  {Ong}, \citenamefont {Cava},\ and\ \citenamefont
  {Bernevig}}]{58wang2016time}%
  \BibitemOpen
  \bibfield  {author} {\bibinfo {author} {\bibfnamefont {Z.}~\bibnamefont
  {Wang}}, \bibinfo {author} {\bibfnamefont {M.}~\bibnamefont {Vergniory}},
  \bibinfo {author} {\bibfnamefont {S.}~\bibnamefont {Kushwaha}}, \bibinfo
  {author} {\bibfnamefont {M.}~\bibnamefont {Hirschberger}}, \bibinfo {author}
  {\bibfnamefont {E.}~\bibnamefont {Chulkov}}, \bibinfo {author} {\bibfnamefont
  {A.}~\bibnamefont {Ernst}}, \bibinfo {author} {\bibfnamefont
  {N.}~\bibnamefont {Ong}}, \bibinfo {author} {\bibfnamefont {R.~J.}\
  \bibnamefont {Cava}}, \ and\ \bibinfo {author} {\bibfnamefont {B.~A.}\
  \bibnamefont {Bernevig}},\ }\href@noop {} {\bibfield  {journal} {\bibinfo
  {journal} {Phys. Rev. Lett.}\ }\textbf {\bibinfo {volume} {117}},\ \bibinfo
  {pages} {236401} (\bibinfo {year} {2016})}\BibitemShut {NoStop}%
\bibitem [{\citenamefont {Gresch}\ \emph {et~al.}(2018)\citenamefont {Gresch},
  \citenamefont {Wu}, \citenamefont {Winkler}, \citenamefont {H{\"a}uselmann},
  \citenamefont {Troyer},\ and\ \citenamefont
  {Soluyanov}}]{59gresch2018automated}%
  \BibitemOpen
  \bibfield  {author} {\bibinfo {author} {\bibfnamefont {D.}~\bibnamefont
  {Gresch}}, \bibinfo {author} {\bibfnamefont {Q.}~\bibnamefont {Wu}}, \bibinfo
  {author} {\bibfnamefont {G.~W.}\ \bibnamefont {Winkler}}, \bibinfo {author}
  {\bibfnamefont {R.}~\bibnamefont {H{\"a}uselmann}}, \bibinfo {author}
  {\bibfnamefont {M.}~\bibnamefont {Troyer}}, \ and\ \bibinfo {author}
  {\bibfnamefont {A.~A.}\ \bibnamefont {Soluyanov}},\ }\href@noop {} {\bibfield
   {journal} {\bibinfo  {journal} {Phys. Rev. Mater.}\ }\textbf {\bibinfo
  {volume} {2}},\ \bibinfo {pages} {103805} (\bibinfo {year}
  {2018})}\BibitemShut {NoStop}%
\bibitem [{\citenamefont {Liu}\ \emph {et~al.}(2013)\citenamefont {Liu},
  \citenamefont {Shan}, \citenamefont {Yao}, \citenamefont {Yao},\ and\
  \citenamefont {Xiao}}]{60liu2013three}%
  \BibitemOpen
  \bibfield  {author} {\bibinfo {author} {\bibfnamefont {G.-B.}\ \bibnamefont
  {Liu}}, \bibinfo {author} {\bibfnamefont {W.-Y.}\ \bibnamefont {Shan}},
  \bibinfo {author} {\bibfnamefont {Y.}~\bibnamefont {Yao}}, \bibinfo {author}
  {\bibfnamefont {W.}~\bibnamefont {Yao}}, \ and\ \bibinfo {author}
  {\bibfnamefont {D.}~\bibnamefont {Xiao}},\ }\href@noop {} {\bibfield
  {journal} {\bibinfo  {journal} {Phys. Rev. B}\ }\textbf {\bibinfo {volume}
  {88}},\ \bibinfo {pages} {085433} (\bibinfo {year} {2013})}\BibitemShut
  {NoStop}%
\bibitem [{\citenamefont {Huang}\ \emph {et~al.}(2016)\citenamefont {Huang},
  \citenamefont {Liu}, \citenamefont {Vanderbilt},\ and\ \citenamefont
  {Duan}}]{61huang2016topological}%
  \BibitemOpen
  \bibfield  {author} {\bibinfo {author} {\bibfnamefont {H.}~\bibnamefont
  {Huang}}, \bibinfo {author} {\bibfnamefont {J.}~\bibnamefont {Liu}}, \bibinfo
  {author} {\bibfnamefont {D.}~\bibnamefont {Vanderbilt}}, \ and\ \bibinfo
  {author} {\bibfnamefont {W.}~\bibnamefont {Duan}},\ }\href@noop {} {\bibfield
   {journal} {\bibinfo  {journal} {Phys. Rev. B}\ }\textbf {\bibinfo {volume}
  {93}},\ \bibinfo {pages} {201114} (\bibinfo {year} {2016})}\BibitemShut
  {NoStop}%
\bibitem [{\citenamefont {Halilov}\ \emph {et~al.}(1998)\citenamefont
  {Halilov}, \citenamefont {Perlov}, \citenamefont {Oppeneer}, \citenamefont
  {Yaresko},\ and\ \citenamefont {Antonov}}]{62halilov1998magnetocrystalline}%
  \BibitemOpen
  \bibfield  {author} {\bibinfo {author} {\bibfnamefont {S.}~\bibnamefont
  {Halilov}}, \bibinfo {author} {\bibfnamefont {A.~Y.}\ \bibnamefont {Perlov}},
  \bibinfo {author} {\bibfnamefont {P.}~\bibnamefont {Oppeneer}}, \bibinfo
  {author} {\bibfnamefont {A.}~\bibnamefont {Yaresko}}, \ and\ \bibinfo
  {author} {\bibfnamefont {V.}~\bibnamefont {Antonov}},\ }\href@noop {}
  {\bibfield  {journal} {\bibinfo  {journal} {Phys. Rev. B}\ }\textbf {\bibinfo
  {volume} {57}},\ \bibinfo {pages} {9557} (\bibinfo {year}
  {1998})}\BibitemShut {NoStop}%
\bibitem [{\citenamefont {Kopnin}\ \emph {et~al.}(2011)\citenamefont {Kopnin},
  \citenamefont {Heikkil{\"a}},\ and\ \citenamefont
  {Volovik}}]{63kopnin2011high}%
  \BibitemOpen
  \bibfield  {author} {\bibinfo {author} {\bibfnamefont {N.}~\bibnamefont
  {Kopnin}}, \bibinfo {author} {\bibfnamefont {T.}~\bibnamefont
  {Heikkil{\"a}}}, \ and\ \bibinfo {author} {\bibfnamefont {G.}~\bibnamefont
  {Volovik}},\ }\href@noop {} {\bibfield  {journal} {\bibinfo  {journal} {Phys.
  Rev. B}\ }\textbf {\bibinfo {volume} {83}},\ \bibinfo {pages} {220503}
  (\bibinfo {year} {2011})}\BibitemShut {NoStop}%
\bibitem [{\citenamefont {Volovik}(2015)}]{64volovik2015standard}%
  \BibitemOpen
  \bibfield  {author} {\bibinfo {author} {\bibfnamefont {G.}~\bibnamefont
  {Volovik}},\ }\href@noop {} {\bibfield  {journal} {\bibinfo  {journal} {Phys.
  Scripta}\ }\textbf {\bibinfo {volume} {2015}},\ \bibinfo {pages} {014014}
  (\bibinfo {year} {2015})}\BibitemShut {NoStop}%
\bibitem [{\citenamefont {Heikkil{\"a}}\ and\ \citenamefont
  {Volovik}(2016)}]{65heikkila2016flat}%
  \BibitemOpen
  \bibfield  {author} {\bibinfo {author} {\bibfnamefont {T.~T.}\ \bibnamefont
  {Heikkil{\"a}}}\ and\ \bibinfo {author} {\bibfnamefont {G.~E.}\ \bibnamefont
  {Volovik}},\ }in\ \href@noop {} {\emph {\bibinfo {booktitle} {Basic Physics
  of Functionalized Graphite}}}\ (\bibinfo  {publisher} {Springer},\ \bibinfo
  {year} {2016})\ pp.\ \bibinfo {pages} {123--143}\BibitemShut {NoStop}%
\bibitem [{\citenamefont {Chang}\ \emph {et~al.}(2016)\citenamefont {Chang},
  \citenamefont {Xu}, \citenamefont {Zheng}, \citenamefont {Singh},
  \citenamefont {Hsu}, \citenamefont {Bian}, \citenamefont {Alidoust},
  \citenamefont {Belopolski}, \citenamefont {Sanchez}, \citenamefont {Zhang}
  \emph {et~al.}}]{66chang2016room}%
  \BibitemOpen
  \bibfield  {author} {\bibinfo {author} {\bibfnamefont {G.}~\bibnamefont
  {Chang}}, \bibinfo {author} {\bibfnamefont {S.-Y.}\ \bibnamefont {Xu}},
  \bibinfo {author} {\bibfnamefont {H.}~\bibnamefont {Zheng}}, \bibinfo
  {author} {\bibfnamefont {B.}~\bibnamefont {Singh}}, \bibinfo {author}
  {\bibfnamefont {C.-H.}\ \bibnamefont {Hsu}}, \bibinfo {author} {\bibfnamefont
  {G.}~\bibnamefont {Bian}}, \bibinfo {author} {\bibfnamefont {N.}~\bibnamefont
  {Alidoust}}, \bibinfo {author} {\bibfnamefont {I.}~\bibnamefont
  {Belopolski}}, \bibinfo {author} {\bibfnamefont {D.~S.}\ \bibnamefont
  {Sanchez}}, \bibinfo {author} {\bibfnamefont {S.}~\bibnamefont {Zhang}},
  \emph {et~al.},\ }\href@noop {} {\bibfield  {journal} {\bibinfo  {journal}
  {Sci. Rep.}\ }\textbf {\bibinfo {volume} {6}},\ \bibinfo {pages} {38839}
  (\bibinfo {year} {2016})}\BibitemShut {NoStop}%
\bibitem [{\citenamefont {Chang}\ \emph
  {et~al.}(2017{\natexlab{b}})\citenamefont {Chang}, \citenamefont {Xu},
  \citenamefont {Zhou}, \citenamefont {Huang}, \citenamefont {Singh},
  \citenamefont {Wang}, \citenamefont {Belopolski}, \citenamefont {Yin},
  \citenamefont {Zhang}, \citenamefont {Bansil} \emph
  {et~al.}}]{67chang2017topological}%
  \BibitemOpen
  \bibfield  {author} {\bibinfo {author} {\bibfnamefont {G.}~\bibnamefont
  {Chang}}, \bibinfo {author} {\bibfnamefont {S.-Y.}\ \bibnamefont {Xu}},
  \bibinfo {author} {\bibfnamefont {X.}~\bibnamefont {Zhou}}, \bibinfo {author}
  {\bibfnamefont {S.-M.}\ \bibnamefont {Huang}}, \bibinfo {author}
  {\bibfnamefont {B.}~\bibnamefont {Singh}}, \bibinfo {author} {\bibfnamefont
  {B.}~\bibnamefont {Wang}}, \bibinfo {author} {\bibfnamefont {I.}~\bibnamefont
  {Belopolski}}, \bibinfo {author} {\bibfnamefont {J.}~\bibnamefont {Yin}},
  \bibinfo {author} {\bibfnamefont {S.}~\bibnamefont {Zhang}}, \bibinfo
  {author} {\bibfnamefont {A.}~\bibnamefont {Bansil}},  \emph {et~al.},\
  }\href@noop {} {\bibfield  {journal} {\bibinfo  {journal} {Phys. Rev. Lett.}\
  }\textbf {\bibinfo {volume} {119}},\ \bibinfo {pages} {156401} (\bibinfo
  {year} {2017}{\natexlab{b}})}\BibitemShut {NoStop}%
\bibitem [{\citenamefont {Xu}\ \emph {et~al.}(2018)\citenamefont {Xu},
  \citenamefont {Liu}, \citenamefont {Shi}, \citenamefont {Muechler},
  \citenamefont {Gayles}, \citenamefont {Felser},\ and\ \citenamefont
  {Sun}}]{68xu2018topological}%
  \BibitemOpen
  \bibfield  {author} {\bibinfo {author} {\bibfnamefont {Q.}~\bibnamefont
  {Xu}}, \bibinfo {author} {\bibfnamefont {E.}~\bibnamefont {Liu}}, \bibinfo
  {author} {\bibfnamefont {W.}~\bibnamefont {Shi}}, \bibinfo {author}
  {\bibfnamefont {L.}~\bibnamefont {Muechler}}, \bibinfo {author}
  {\bibfnamefont {J.}~\bibnamefont {Gayles}}, \bibinfo {author} {\bibfnamefont
  {C.}~\bibnamefont {Felser}}, \ and\ \bibinfo {author} {\bibfnamefont
  {Y.}~\bibnamefont {Sun}},\ }\href@noop {} {\bibfield  {journal} {\bibinfo
  {journal} {Phys. Rev. B}\ }\textbf {\bibinfo {volume} {97}},\ \bibinfo
  {pages} {235416} (\bibinfo {year} {2018})}\BibitemShut {NoStop}%
\bibitem [{\citenamefont {Noky}\ \emph {et~al.}(2019)\citenamefont {Noky},
  \citenamefont {Xu}, \citenamefont {Felser},\ and\ \citenamefont
  {Sun}}]{69noky2019large}%
  \BibitemOpen
  \bibfield  {author} {\bibinfo {author} {\bibfnamefont {J.}~\bibnamefont
  {Noky}}, \bibinfo {author} {\bibfnamefont {Q.}~\bibnamefont {Xu}}, \bibinfo
  {author} {\bibfnamefont {C.}~\bibnamefont {Felser}}, \ and\ \bibinfo {author}
  {\bibfnamefont {Y.}~\bibnamefont {Sun}},\ }\href@noop {} {\bibfield
  {journal} {\bibinfo  {journal} {Phys. Rev. B}\ }\textbf {\bibinfo {volume}
  {99}},\ \bibinfo {pages} {165117} (\bibinfo {year} {2019})}\BibitemShut
  {NoStop}%
\bibitem [{\citenamefont {Belopolski}\ \emph {et~al.}(2019)\citenamefont
  {Belopolski}, \citenamefont {Manna}, \citenamefont {Sanchez}, \citenamefont
  {Chang}, \citenamefont {Ernst}, \citenamefont {Yin}, \citenamefont {Zhang},
  \citenamefont {Cochran}, \citenamefont {Shumiya}, \citenamefont {Zheng} \emph
  {et~al.}}]{70belopolski2019discovery}%
  \BibitemOpen
  \bibfield  {author} {\bibinfo {author} {\bibfnamefont {I.}~\bibnamefont
  {Belopolski}}, \bibinfo {author} {\bibfnamefont {K.}~\bibnamefont {Manna}},
  \bibinfo {author} {\bibfnamefont {D.~S.}\ \bibnamefont {Sanchez}}, \bibinfo
  {author} {\bibfnamefont {G.}~\bibnamefont {Chang}}, \bibinfo {author}
  {\bibfnamefont {B.}~\bibnamefont {Ernst}}, \bibinfo {author} {\bibfnamefont
  {J.}~\bibnamefont {Yin}}, \bibinfo {author} {\bibfnamefont {S.~S.}\
  \bibnamefont {Zhang}}, \bibinfo {author} {\bibfnamefont {T.}~\bibnamefont
  {Cochran}}, \bibinfo {author} {\bibfnamefont {N.}~\bibnamefont {Shumiya}},
  \bibinfo {author} {\bibfnamefont {H.}~\bibnamefont {Zheng}},  \emph
  {et~al.},\ }\href@noop {} {\bibfield  {journal} {\bibinfo  {journal}
  {Science}\ }\textbf {\bibinfo {volume} {365}},\ \bibinfo {pages} {1278}
  (\bibinfo {year} {2019})}\BibitemShut {NoStop}%
\bibitem [{\citenamefont {Liu}\ \emph {et~al.}(2019)\citenamefont {Liu},
  \citenamefont {Liang}, \citenamefont {Liu}, \citenamefont {Xu}, \citenamefont
  {Li}, \citenamefont {Chen}, \citenamefont {Pei}, \citenamefont {Shi},
  \citenamefont {Mo}, \citenamefont {Dudin} \emph
  {et~al.}}]{71liu2019magnetic}%
  \BibitemOpen
  \bibfield  {author} {\bibinfo {author} {\bibfnamefont {D.}~\bibnamefont
  {Liu}}, \bibinfo {author} {\bibfnamefont {A.}~\bibnamefont {Liang}}, \bibinfo
  {author} {\bibfnamefont {E.}~\bibnamefont {Liu}}, \bibinfo {author}
  {\bibfnamefont {Q.}~\bibnamefont {Xu}}, \bibinfo {author} {\bibfnamefont
  {Y.}~\bibnamefont {Li}}, \bibinfo {author} {\bibfnamefont {C.}~\bibnamefont
  {Chen}}, \bibinfo {author} {\bibfnamefont {D.}~\bibnamefont {Pei}}, \bibinfo
  {author} {\bibfnamefont {W.}~\bibnamefont {Shi}}, \bibinfo {author}
  {\bibfnamefont {S.}~\bibnamefont {Mo}}, \bibinfo {author} {\bibfnamefont
  {P.}~\bibnamefont {Dudin}},  \emph {et~al.},\ }\href@noop {} {\bibfield
  {journal} {\bibinfo  {journal} {Science}\ }\textbf {\bibinfo {volume}
  {365}},\ \bibinfo {pages} {1282} (\bibinfo {year} {2019})}\BibitemShut
  {NoStop}%
\bibitem [{\citenamefont {Morali}\ \emph {et~al.}(2019)\citenamefont {Morali},
  \citenamefont {Batabyal}, \citenamefont {Nag}, \citenamefont {Liu},
  \citenamefont {Xu}, \citenamefont {Sun}, \citenamefont {Yan}, \citenamefont
  {Felser}, \citenamefont {Avraham},\ and\ \citenamefont
  {Beidenkopf}}]{72Morali1286}%
  \BibitemOpen
  \bibfield  {author} {\bibinfo {author} {\bibfnamefont {N.}~\bibnamefont
  {Morali}}, \bibinfo {author} {\bibfnamefont {R.}~\bibnamefont {Batabyal}},
  \bibinfo {author} {\bibfnamefont {P.~K.}\ \bibnamefont {Nag}}, \bibinfo
  {author} {\bibfnamefont {E.}~\bibnamefont {Liu}}, \bibinfo {author}
  {\bibfnamefont {Q.}~\bibnamefont {Xu}}, \bibinfo {author} {\bibfnamefont
  {Y.}~\bibnamefont {Sun}}, \bibinfo {author} {\bibfnamefont {B.}~\bibnamefont
  {Yan}}, \bibinfo {author} {\bibfnamefont {C.}~\bibnamefont {Felser}},
  \bibinfo {author} {\bibfnamefont {N.}~\bibnamefont {Avraham}}, \ and\
  \bibinfo {author} {\bibfnamefont {H.}~\bibnamefont {Beidenkopf}},\
  }\href@noop {} {\bibfield  {journal} {\bibinfo  {journal} {Science}\ }\textbf
  {\bibinfo {volume} {365}},\ \bibinfo {pages} {1286} (\bibinfo {year}
  {2019})}\BibitemShut {NoStop}%
\bibitem [{\citenamefont {Chen}\ \emph {et~al.}(2011)\citenamefont {Chen},
  \citenamefont {Zhu}, \citenamefont {Xiao},\ and\ \citenamefont
  {Zhang}}]{73chen2011co}%
  \BibitemOpen
  \bibfield  {author} {\bibinfo {author} {\bibfnamefont {H.}~\bibnamefont
  {Chen}}, \bibinfo {author} {\bibfnamefont {W.}~\bibnamefont {Zhu}}, \bibinfo
  {author} {\bibfnamefont {D.}~\bibnamefont {Xiao}}, \ and\ \bibinfo {author}
  {\bibfnamefont {Z.}~\bibnamefont {Zhang}},\ }\href@noop {} {\bibfield
  {journal} {\bibinfo  {journal} {Phys. Rev. Lett.}\ }\textbf {\bibinfo
  {volume} {107}},\ \bibinfo {pages} {056804} (\bibinfo {year}
  {2011})}\BibitemShut {NoStop}%
\bibitem [{\citenamefont {Rajamathi}\ \emph {et~al.}(2017)\citenamefont
  {Rajamathi}, \citenamefont {Gupta}, \citenamefont {Kumar}, \citenamefont
  {Yang}, \citenamefont {Sun}, \citenamefont {S{\"u}{\ss}}, \citenamefont
  {Shekhar}, \citenamefont {Schmidt}, \citenamefont {Blumtritt}, \citenamefont
  {Werner} \emph {et~al.}}]{74rajamathi2017weyl}%
  \BibitemOpen
  \bibfield  {author} {\bibinfo {author} {\bibfnamefont {C.~R.}\ \bibnamefont
  {Rajamathi}}, \bibinfo {author} {\bibfnamefont {U.}~\bibnamefont {Gupta}},
  \bibinfo {author} {\bibfnamefont {N.}~\bibnamefont {Kumar}}, \bibinfo
  {author} {\bibfnamefont {H.}~\bibnamefont {Yang}}, \bibinfo {author}
  {\bibfnamefont {Y.}~\bibnamefont {Sun}}, \bibinfo {author} {\bibfnamefont
  {V.}~\bibnamefont {S{\"u}{\ss}}}, \bibinfo {author} {\bibfnamefont
  {C.}~\bibnamefont {Shekhar}}, \bibinfo {author} {\bibfnamefont
  {M.}~\bibnamefont {Schmidt}}, \bibinfo {author} {\bibfnamefont
  {H.}~\bibnamefont {Blumtritt}}, \bibinfo {author} {\bibfnamefont
  {P.}~\bibnamefont {Werner}},  \emph {et~al.},\ }\href@noop {} {\bibfield
  {journal} {\bibinfo  {journal} {Adv. Mater.}\ }\textbf {\bibinfo {volume}
  {29}},\ \bibinfo {pages} {1606202} (\bibinfo {year} {2017})}\BibitemShut
  {NoStop}%
\bibitem [{\citenamefont {Li}\ \emph {et~al.}(2018)\citenamefont {Li},
  \citenamefont {Ma}, \citenamefont {Xie}, \citenamefont {Feng}, \citenamefont
  {Ullah}, \citenamefont {Li}, \citenamefont {Dong}, \citenamefont {Li},
  \citenamefont {Li},\ and\ \citenamefont {Chen}}]{75li2018topological}%
  \BibitemOpen
  \bibfield  {author} {\bibinfo {author} {\bibfnamefont {J.}~\bibnamefont
  {Li}}, \bibinfo {author} {\bibfnamefont {H.}~\bibnamefont {Ma}}, \bibinfo
  {author} {\bibfnamefont {Q.}~\bibnamefont {Xie}}, \bibinfo {author}
  {\bibfnamefont {S.}~\bibnamefont {Feng}}, \bibinfo {author} {\bibfnamefont
  {S.}~\bibnamefont {Ullah}}, \bibinfo {author} {\bibfnamefont
  {R.}~\bibnamefont {Li}}, \bibinfo {author} {\bibfnamefont {J.}~\bibnamefont
  {Dong}}, \bibinfo {author} {\bibfnamefont {D.}~\bibnamefont {Li}}, \bibinfo
  {author} {\bibfnamefont {Y.}~\bibnamefont {Li}}, \ and\ \bibinfo {author}
  {\bibfnamefont {X.-Q.}\ \bibnamefont {Chen}},\ }\href@noop {} {\bibfield
  {journal} {\bibinfo  {journal} {Sci. China Mater.}\ }\textbf {\bibinfo
  {volume} {61}},\ \bibinfo {pages} {23} (\bibinfo {year} {2018})}\BibitemShut
  {NoStop}%
\end{thebibliography}%

\end{document}